\documentclass[trackchanges]{aastex701}
\usepackage{amsmath}
\usepackage{threeparttable}

\begin{document}

\title{FAST Polarization Catalog of FRB 20240114A}

\author[orcid=0000-0002-9332-5562,sname='Wang']{Tian-Cong Wang}
\altaffiliation{These authors contributed equally to this work.}
\affiliation{School of Physics and Astronomy, Beijing Normal University, Beijing 100875, China}
\affiliation{Institute for Frontiers in Astronomy and Astrophysics, Beijing Normal University, Beijing 102206, China}
\email{wangtc@mail.bnu.edu.cn} 

\author[orcid=0009-0005-8586-3001,sname='Zhang']{Jun-Shuo Zhang}
\altaffiliation{These authors contributed equally to this work.}
\affiliation{National Astronomical Observatories, Chinese Academy of Sciences, Beijing 100101, China}
\affiliation{University of Chinese Academy of Sciences, Beijing 100049, China}
\email{zhangjs@bao.ac.cn} 

\author[orcid=0000-0002-2552-7277,sname='Liu']{Xiao-Hui Liu}
\altaffiliation{These authors contributed equally to this work.}
\affiliation{National Astronomical Observatories, Chinese Academy of Sciences, Beijing 100101, China}
\affiliation{University of Chinese Academy of Sciences, Beijing 100049, China}
\email{liuxh@bao.ac.cn}

\author[orcid=0000-0001-9036-8543,sname='Wang']{Wei-Yang Wang}
\affiliation{University of Chinese Academy of Sciences, Beijing 100049, China}
\email{wywang@ucas.ac.cn} 

\author[orcid=0000-0002-3386-7159,sname='Wang']{Pei Wang}
\affiliation{National Astronomical Observatories, Chinese Academy of Sciences, Beijing 100101, China}
\affiliation{Institute for Frontiers in Astronomy and Astrophysics, Beijing Normal University, Beijing 102206, China}
\affiliation{State Key Laboratory of Radio Astronomy and Technology, Beijing 100101, China}
\email[show]{wangpei@nao.cas.cn} 

\author[orcid=0000-0003-2516-6288,sname='Gao']{He Gao}
\affiliation{School of Physics and Astronomy, Beijing Normal University, Beijing 100875, China}
\affiliation{Institute for Frontiers in Astronomy and Astrophysics, Beijing Normal University, Beijing 102206, China}
\email[show]{gaohe@bnu.edu.cn} 

\author[orcid=0000-0003-3010-7661, sname='Li']{Di Li}
\affiliation{New Cornerstone Science Laboratory, Department of Astronomy, Tsinghua University, Beijing 100084, China}
\affiliation{National Astronomical Observatories, Chinese Academy of Sciences, Beijing 100101, China}
\affiliation{State Key Laboratory of Radio Astronomy and Technology, Beijing 100101, China}
\email[show]{dili@tsinghua.edu.cn}

\author[orcid=0000-0002-9725-2524,sname='Zhang']{Bing Zhang}
\affiliation{The Hong Kong Institute for Astronomy and Astrophysics, the University of Hong Kong, Pokfulam, Hong Kong, China}
\affiliation{Department of Physics, the University of Hong Kong, Pokfulam, Hong Kong, China}
\affiliation{Nevada Center for Astrophysics, University of Nevada, Las Vegas, NV 89154, USA}
\affiliation{Department of Physics and Astronomy, University of Nevada Las Vegas, Las Vegas, NV 89154, USA}
\email[show]{bzhang1@hku.hk} 

\author[orcid=0000-0001-5105-4058,sname='Zhu']{Wei-Wei Zhu}
\affiliation{National Astronomical Observatories, Chinese Academy of Sciences, Beijing 100101, China}
\affiliation{Institute for Frontiers in Astronomy and Astrophysics, Beijing Normal University, Beijing 102206, China}
\affiliation{State Key Laboratory of Radio Astronomy and Technology, Beijing 100101, China}
\email{zhuww@nao.cas.cn} 

\author{Jin-Lin Han}
\affiliation{National Astronomical Observatories, Chinese Academy of Sciences, Beijing 100101, China}
\affiliation{University of Chinese Academy of Sciences, Beijing 100049, China}
\affiliation{State Key Laboratory of Radio Astronomy and Technology, Beijing 100101, China}
\email{hjl@nao.cas.cn}

\author{Ke-Jia Lee}
\affiliation{Department of Astronomy, School of physics, Peking University, Beijing 100871, China}
\affiliation{National Astronomical Observatories, Chinese Academy of Sciences, Beijing 100101, China}
\affiliation{Yunnan Astronomical Observatories, Chinese Academy of Sciences, Kunming 650216, China}
\affiliation{Beijing Laser Acceleration Innovation Center, Huairou, Beijing 101400, China}
\email{kjlee@pku.edu.cn}

\author{Ye Li}
\affiliation{Purple Mountain Observatory, Chinese Academy of Sciences, Nanjing 210023, China}
\email{yeli@pmo.ac.cn}

\author[0000-0002-7420-9988]{Dengke Zhou}
\affiliation{Research Center for Astronomical Computing, Zhejiang Laboratory, Hangzhou 311121, China}
\email{zdk@zhejianglab.com}

\author{Wan-Jin Lu}
\affiliation{National Astronomical Observatories, Chinese Academy of Sciences, Beijing 100101, China}
\affiliation{University of Chinese Academy of Sciences, Beijing 100049, China}
\email{wjlu@bao.ac.cn}

\author[0000-0001-5649-2591]{Jintao Xie}
\affiliation{School of Computer Science and Engineering, Sichuan University of Science and Engineering, Yibin 644000, China}
\email{xiejintao@suse.edu.cn}

\author[0000-0001-9956-6298]{Jianhua Fang}
\affiliation{Research Center for Astronomical Computing, Zhejiang Laboratory, Hangzhou 311121, China}
\email{fangjh@zhejianglab.com}

\author{Jin-Huang Cao}
\affiliation{National Astronomical Observatories, Chinese Academy of Sciences, Beijing 100101, China}
\affiliation{University of Chinese Academy of Sciences, Beijing 100049, China}
\email{caojinhuang22@mails.ucas.ac.cn}

\author{Chen-Chen Miao}
\affiliation{College of Physics and Electronic Engineering, Qilu Normal University, Jinan 250200, China}
\email{miaocc@nao.cas.cn}

\author{Yu-Hao Zhu}
\affiliation{National Astronomical Observatories, Chinese Academy of Sciences, Beijing 100101, China}
\affiliation{University of Chinese Academy of Sciences, Beijing 100049, China}
\email{zhuyh@bao.ac.cn}

\author[0009-0000-4795-8767]{Yunchuan Chen}
\affiliation{Research Center for Astronomical Computing, Zhejiang Laboratory, Hangzhou 311121, China}
\email{chenych@zhejianglab.org}

\author{Si-Lu Xu}
\affiliation{National Astronomical Observatories, Chinese Academy of Sciences, Beijing 100101, China}
\affiliation{University of Chinese Academy of Sciences, Beijing 100049, China}
\email{xusilu25@mails.ucas.ac.cn}

\author[0009-0000-6108-2730]{Huaxi Chen}
\affiliation{Research Center for Astronomical Computing, Zhejiang Laboratory, Hangzhou 311121, China}
\email{chenhuaxi@zhejianglab.org}

\author{Xiao-Feng Cheng}
\affiliation{Research Center for Astronomical Computing, Zhejiang Laboratory, Hangzhou 311121, China}
\email{chengxfmail@163.com}

\author[0000-0001-6021-5933]{Qin Wu}
\affiliation{School of Astronomy and Space Science, Nanjing University, Nanjing 210023, China}
\email{wqin@smail.nju.edu.cn}

\author{Shuo Cao}
\affiliation{Yunnan Astronomical Observatories, Chinese Academy of Sciences, Kunming 650216, China}
\affiliation{University of Chinese Academy of Sciences, Beijing 100049, China}
\email{caoshuo@ynao.ac.cn}

\author[0009-0002-3020-9123]{Long-Xuan Zhang}
\affiliation{School of Physics, Huazhong University of Science and Technology, Wuhan, 430074. China}
\email{d202580060@hust.edu.cn}

\author{Shi-Yan Tian}
\affiliation{School of Physics, Huazhong University of Science and Technology, Wuhan, 430074. China}
\email{2226247263@qq.com}

\author{Yong-Kun Zhang}
\affiliation{National Astronomical Observatories, Chinese Academy of Sciences, Beijing 100101, China}
\email{ykzhang@nao.cas.cn}

\author{Yi Feng}
\affiliation{Research Center for Astronomical Computing, Zhejiang Laboratory, Hangzhou 311121, China}
\affiliation{Institute for Astronomy, School of Physics, Zhejiang University, Hangzhou 310027, China}
\email{yifeng@zhejianglab.org}

\author[0000-0002-6423-6106]{De-Jiang Zhou}
\affiliation{National Astronomical Observatories, Chinese Academy of Sciences, Beijing 100101, China}
\email{djzhou@nao.cas.cn}

\author{Jia-Rui Niu}
\affiliation{National Astronomical Observatories, Chinese Academy of Sciences, Beijing 100101, China}
\email{niujiarui@nao.cas.cn}

\author{Heng Xu}
\affiliation{National Astronomical Observatories, Chinese Academy of Sciences, Beijing 100101, China}
\email{hengxu@pku.edu.cn}

\author[orcid=0000-0001-6475-8863,sname='Chen']{Xuelei Chen}
\affiliation{State Key Laboratory of Radio Astronomy and Technology, Beijing 100101, China}
\affiliation{University of Chinese Academy of Sciences, Beijing 100049, China}
\email{xuelei@cosmology.bao.ac.cn}

\author[0000-0001-6374-8313]{Yuan-Pei Yang}
\affiliation{South-Western Institute for Astronomy Research, Key Laboratory of Survey Science of Yunnan Province, Yunnan University, Kunming 650500, China}
\email{ypyang@ynu.edu.cn}

\author{Dong-Zi Li}
\affiliation{Department of Astrophysical Sciences, Princeton University, Princeton, NJ 89154, USA}
\affiliation{Department of Astronomy, Tsinghua University, Beijing 100084, China.}
\email{dzli@tsinghua.edu.cn}

\author{Fa-Yin Wang}
\affiliation{School of Astronomy and Space Science, Nanjing University, Nanjing 210023, China}
\affiliation{Key Laboratory of Modern Astronomy and Astrophysics (Nanjing University), Ministry of Education, Nanjing 210093, China}
\email{fayinwang@nju.edu.cn}

\author[0000-0002-9390-9672]{Chao-Wei Tsai}
\affiliation{National Astronomical Observatories, Chinese Academy of Sciences, Beijing 100101, China}
\affiliation{Institute for Frontiers in Astronomy and Astrophysics, Beijing Normal University,  Beijing 102206, China}
\affiliation{University of Chinese Academy of Sciences, Beijing 100049, China}
\affiliation{State Key Laboratory of Radio Astronomy and Technology, Beijing 100101, China}
\email{cwtsai@nao.cas.cn}

\author{Wen-Fei Yu}
\affiliation{Shanghai Astronomical Observatory, Chinese Academy of Sciences, Shanghai 200030, China}
\email{wenfei@shao.ac.cn}

\author[0000-0001-6651-7799]{Chen-Hui Niu}
\affiliation{Institute of Astrophysics, Central China Normal University, Wuhan 430079, China}
\email{niuchenhui@ccnu.edu.cn}

\author[0000-0002-9642-9682]{Jia-Wei Luo}
\affiliation{College of Physics and Hebei Key Laboratory of Photophysics Research and Application, Hebei Normal University, Shijiazhuang 050024, China}
\affiliation{Shijiazhuang Key Laboratory of Astronomy and Space Science, Hebei Normal University, Shijiazhuang 050024, China}
\email{ljw@hebtu.edu.cn}

\author[0000-0002-4300-121X]{Rui Luo}
\affiliation{Department of Astronomy, School of Physics and Materials Science, Guangzhou University, Guangzhou 510006, China}
\email{rui.luo@gzhu.edu.cn}

\author{E. G\"{u}gercino\u{g}lu}
\affiliation{National Astronomical Observatories, Chinese Academy of Sciences, Beijing 100101, China}
\affiliation{School of Arts and Sciences, Qingdao Binhai University, Qingdao 266555, China}
\email{egugercinoglu@gmail.com}

\author{Zi-Wei Wu}
\affiliation{National Astronomical Observatories, Chinese Academy of Sciences, Beijing 100101, China}
\email{wuzw@bao.ac.cn}

\author[0000-0002-4327-711X]{Chun-Feng Zhang}
\affiliation{National Astronomical Observatories, Chinese Academy of Sciences, Beijing 100101, China}
\email{cf.zhang@pku.edu.cn}

\author{Xiang-Lei Chen}
\affiliation{National Astronomical Observatories, Chinese Academy of Sciences, Beijing 100101, China}
\email{xiangleichen@nao.cas.cn}

\author[orcid=0009-0006-0803-0505,sname='Feng']{Shuai Feng}
\affiliation{National Astronomical Observatories, Chinese Academy of Sciences, Beijing 100101, China}
\affiliation{University of Chinese Academy of Sciences, Beijing 100049, China}
\email{fshuai@bao.ac.cn}

\author[0000-0002-6165-0977]{Xiang-Han Cui}
\affiliation{National Astronomical Observatories, Chinese Academy of Sciences, Beijing 100101, China}
\affiliation{The Netherlands Institute for Radio Astronomy, Oude Hoogeveensedijk 4,7991 PD Dwingeloo, The Netherlands}
\email{cuixianghan@nao.cas.cn}

\author{Qing-Yue Qu}
\affiliation{National Astronomical Observatories, Chinese Academy of Sciences, Beijing 100101, China}
\affiliation{University of Chinese Academy of Sciences, Beijing 100049, China}
\email{quqy@bao.ac.cn}

\author[0000-0003-4721-4869]{Yuan-Hong Qu}
\affiliation{Nevada Center for Astrophysics, University of Nevada, Las Vegas, NV 89154, USA}
\affiliation{Department of Physics and Astronomy, University of Nevada Las Vegas, Las Vegas, NV 89154, USA}
\email{yuanhong.qu@unlv.edu}

\author[0000-0002-9434-4773]{Bo-Jun Wang}
\affiliation{National Astronomical Observatories, Chinese Academy of Sciences, Beijing 100101, China}
\email{wangbj@bao.ac.cn}

\author{Yi-Dan Wang}
\affiliation{National Astronomical Observatories, Chinese Academy of Sciences, Beijing 100101, China}
\affiliation{University of Chinese Academy of Sciences, Beijing 100049, China}
\email{wangyd@bao.ac.cn}

\author{Lin Lin}
\affiliation{School of Physics and Astronomy, Beijing Normal University, Beijing 100875, China}
\email{llin@bnu.edu.cn}

\author{Ai-Yuan Yang}
\affiliation{National Astronomical Observatories, Chinese Academy of Sciences, Beijing 100101, China}
\affiliation{State Key Laboratory of Radio Astronomy and Technology, Beijing 100101, China}
\email{yangay@bao.ac.cn}

\author[0000-0002-5400-3261]{Yuan-Chuan Zou}
\affiliation{School of Physics, Huazhong University of Science and Technology, Wuhan, 430074. China}
\affiliation{Purple Mountain Observatory, Chinese Academy of Sciences, Nanjing 210023, China}
\email{zouyc@hust.edu.cn}

\author{Yu-Xiang Huang}
\affiliation{Yunnan Astronomical Observatories, Chinese Academy of Sciences, Kunming 650216, China}
\email{huangyuxiang@ynao.ac.cn}

\author[0000-0002-1056-5895]{Wei-Cong Jing}
\affiliation{National Astronomical Observatories, Chinese Academy of Sciences, Beijing 100101, China}
\email{jingweicong@nao.cas.cn}

\author{Jian Li}
\affiliation{Department of Astronomy, University of Science and Technology of China, Hefei 230026, China}
\affiliation{School of Astronomy and Space Science, University of Science and Technology of China, Hefei 230026, China}
\email{jianli@ustc.edu.cn}

\author[0000-0001-7199-2906]{Yong-Feng Huang}
\affiliation{School of Astronomy and Space Science, Nanjing University, Nanjing 210023, China}
\affiliation{Key Laboratory of Modern Astronomy and Astrophysics (Nanjing University), Ministry of Education, Nanjing 210093, China}
\email{hyf@nju.edu.cn}

\author[0000-0001-7746-9462]{Su-Ming Weng}
\affiliation{National Key Laboratory of Dark Matter Physics, School of Physics and Astronomy, Shanghai Jiao Tong University, Shanghai 200240, China}
\affiliation{Laboratory for Laser Plasmas and Collaborative Innovation Centre of IFSA, Shanghai Jiao Tong University, Shanghai 200240, China} 
\email{wengsuming@sjtu.edu.cn}

\author[0000-0002-5799-9869]{Shi-Han Yew}
\affiliation{National Key Laboratory of Dark Matter Physics, School of Physics and Astronomy, Shanghai Jiao Tong University, Shanghai 200240, China}
\affiliation{Laboratory for Laser Plasmas and Collaborative Innovation Centre of IFSA, Shanghai Jiao Tong University, Shanghai 200240, China}
\email{yushihan@sjtu.edu.cn}

\author[0000-0002-6299-1263]{Xue-Feng Wu}
\affiliation{Purple Mountain Observatory, Chinese Academy of Sciences, Nanjing 210023, China}
\email{xfwu@pmo.ac.cn}

\author{Lei Zhang}
\affiliation{National Astronomical Observatories, Chinese Academy of Sciences, Beijing 100101, China}
\affiliation{Centre for Astrophysics and Supercomputing, Swinburne University of Technology, Hawthorn, VIC 3122, Australia}
\email{leizhang996@nao.cas.cn}

\author{Ru-Shuang Zhao}
\affiliation{Guizhou Provincial Key Laboratory of Radio Astronomy and Data Processing, Guizhou Normal University, Guiyang 550001, China}
\email{201907007@gznu.edu.cn}

\correspondingauthor{Pei Wang, He Gao, Li Di, Bing Zhang}


\begin{abstract}

Polarization measurements of fast radio bursts (FRBs) probe the magnetized plasma surrounding their central engines. FRB~20240114A is an exceptionally active repeating source, with 17,356 bursts detected between 2024 January 28 and 2025 May 30 by FAST, enabling studies of the temporal evolution of its polarization properties. In this work, we present a polarimetric catalog of 6,131 bright bursts (with a signal-to-noise ratio S/N $\geq$ 20, 35.3\% of the total sample), including arrival time (MJD$_{\text{topo}}$), dispersion measure (DM), burst width (W$_{\text{eff}}$), bandwidth, Faraday rotation measure (RM), linear and circular polarization degrees ($L/I$, $V/I$), and intrinsic polarization angle (PA$_0$). We confirm a clear temporal evolution of RM: after an initial stable phase, it decreases linearly by $\sim$200 $\rm rad\ m^{-2}$ over 200 days, forming a bimodal distribution, whereas DM remains stable at 529.3 $\pm$ 1.2 $\rm pc\ cm^{-3}$. The linear polarization fraction is generally high, with the 3$\sigma$ lower bound around 76\%, while circular polarization is low, with 1,157 of 17,356 bursts (6.67\%) having $|V|/I$ $\geq$10\%. We perform a power-law fit between $|V|/I$ and $|\textrm{RM}|$, which yields an index of $-2.98 \pm 0.80$. It is found that the combined 2D distribution of $L/I$ versus $V/I$ remains stable, implying that the emission mechanism is largely invariant. Our PA$_0$ measurements show a broad, non-uniform distribution, implying a complex emission geometry. These results suggest that FRB~20240114A resides in a dynamically evolving magneto-ionic environment. This catalog provides a foundation for studies of repeating FRB progenitors and their environments.

\end{abstract}


\section{Introduction}\label{sec:Intro}
Fast radio bursts (FRBs) represent one of the most energetic and enigmatic phenomena in the transient universe \citep{2023RvMP...95c5005Z}. Since their discovery \citep{2007Sci.Lorimer,2013Sci.Thornton}, the identification of repeating sources \citep{2016Natur.Spitler,2019Natur.CHIME/FRB} has provided a crucial path toward understanding their progenitors. Among the various observational tools, radio polarimetry has proven to be exceptionally powerful. Measurements of the linear polarization and the Faraday rotation measure (RM) offer a direct probe into the magneto-ionic environment surrounding the FRB source, potentially revealing the presence of a supernova remnant, a pulsar wind nebula, or a binary companion's wind \citep{2018Natur.Michilli,2020ApJ.YangYP,Wang2022,2023MNRAS.Qu&Zhang}. Furthermore, the temporal evolution of these properties can shed light on the dynamics of this local environment \citep{2021ApJ.Hilmarsson,2022Natur.XuH}.

While individual case studies of bright bursts are valuable, a comprehensive statistical understanding of an FRB's behavior requires large, uniform samples of bursts observed with a single, stable instrument. Such datasets mitigate instrumental biases and allow for the robust investigation of burst-to-burst variations, energy distributions, and long-term temporal evolution.

Recent years have also seen the emergence of systematic FRB polarization catalogs from multiple instruments. For non-repeating FRBs, polarimetric catalogs have been presented for samples detected with DSA-110 \citep{2024ApJ...964..131S}, CHIME/FRB \citep{2024ApJ...968...50P}, and ASKAP \citep{2025PASA...42..133S}. For repeating sources, recent CHIME/FRB studies have established the largest source-level polarization catalogs to date \citep{2023ApJ...951...82M,2025ApJ...982..154N}, enabling comparative studies across repeater populations. In this context, the present work is complementary: rather than a source-level census, we provide a large burst-level, time-resolved polarization catalog for a single exceptionally active repeating FRB, making it possible to investigate burst-to-burst variations and long-term temporal evolution within one source.

The repeating FRB 20240114A, discovered shortly after its initial activity, has established itself as one of the most prolific known sources \citep{2025Shin,2024MNRAS.Tian,2025ApJ.Panda,2025ApJS.Xie,2025Zhangjs,2025arXiv250714708Z,2025Zhanglx}.Its sustained high burst rate over an extended period makes it an ideal target for amassing a large catalog of bursts. The Five-hundred-meter Aperture Spherical radio Telescope (FAST, \citealp{2011IJMPD.Nan}), with its unparalleled sensitivity, is uniquely capable of detecting the faint end of the burst energy distribution with a fluence completeness threshold of 0.026 Jy~ms from such a source \citep{2025Zhangjs}, thereby providing a complete and unbiased sample. This extreme sensitivity also makes FAST a powerful platform for high-precision polarimetry. Previous FAST observations have revealed diverse polarization-angle swings, frequency-dependent polarization behavior, extreme circular polarization, and temporal evolution of the local magneto-ionic environment in repeating FRBs \citep{2020Natur.586..693L,2022Natur.XuH,2022RAA....22l4003J,2023Sci.Anna-Thomas,2024ApJ...972L..20N,2024NSRev..12E.293J,2025ApJ...988...41Z,feng25,2026Sci...391..280L}, demonstrating that polarimetric measurements can serve as a sensitive probe of their extreme and dynamic environments. Building on these results, the present work provides a systematic polarimetric catalog and time-resolved statistical analysis of FRB~20240114A based on a much larger burst sample.

This paper is the first in a series devoted to unveiling the polarization nature of FRB 20240114A through FAST observations. Herein, we focus exclusively on the presentation and public release of the polarization catalog. We describe our 16-month observations from 2024 to 2025, the detailed data reduction pipeline, and the calibration procedures that underpin this dataset. The resulting catalog comprises over 6000 bursts, each with measured time of arrival (TOA), dispersion measure (DM), RM, and linear and circular polarization fractions.

We explicitly state that the comprehensive scientific analysis of this catalog, including the search for periodicity, the detailed modeling of RM evolution, and the investigation of burst sub-populations, will be the focus of subsequent publications in this series. The purpose of this paper is to establish the data product itself, ensuring its reliability and making it available to the broader community for independent analysis.

This paper is organized as follows. In Section \ref{sec:ObsAndData}, we describe the FAST observations and the data processing procedures, including burst search and parameter measurements. The catalog of bursts is presented in Section \ref{sec:CataDesc}, providing a detailed description of the catalog. In Section \ref{sec:quality}, we analyze the data quality and global properties of the bursts, including parameter distributions and correlations. Section \ref{sec:discussion} is dedicated to a discussion of the possible origins of the RM evolution and the complex and variable geometry of the emission region. Finally, a summary and information on data access are provided in Section \ref{sec:summary}.

\section{Observations and Data Processing}\label{sec:ObsAndData}

\subsection{FAST Observations and Burst Search}\label{subsec:FASTobs}

An observation campaign of FRB 20240114A was conducted using the FAST telescope for over sixteen months, from 2024 January 28 to 2025 May 30.  The initial observations employed the position reported by the Canadian Hydrogen Intensity Mapping Experiment (CHIME) \citep{2025Shin} (RA = 21$^\mathrm{h}$27$^\mathrm{m}$39$^\mathrm{s}$.888, Dec = +04$^\circ$21$^\prime$0$^{\prime \prime}$.36). Starting from 15 February 2024 (UT), the pointing was updated to the refined location (RA = 21$^\mathrm{h}$27$^\mathrm{m}$39$^\mathrm{s}$.84, Dec = +04$^\circ$19$^\prime$46$^{\prime \prime}$.3) localized by MeerKAT \citep{2024MNRAS.Tian}. The observing campaign comprised 97 sessions, totaling 57.99 hours of exposure time. Observations covered the 1.0–1.5 GHz band using the central beam of the 19-beam receiver in the tracking mode. Most data were recorded in the 8-bit PSRFITS format, providing full Stokes polarization information, 4096 frequency channels with a frequency resolution of 0.1221 MHz and a time resolution of 49.152 $\mu$s.

We performed single-pulse searches on the FAST data using the PRESTO software package \citep{2001AAS.Ransom}, which employs a matched-filtering technique that applies boxcar functions of varying widths to the dedispersed time series to identify pulses across a range of timescales. The pipeline was described in \cite{2025Wang&Li}. 
To evaluate the completeness of the burst search, we also cross-checked the PRESTO results using two additional search approaches: DRAFTS, a deep-learning-based radio fast transient search pipeline \citep{2025ApJS.ZhangYK}, and the AI-aided pipeline described in \cite{2024WangP}. In the present work, PRESTO serves as the primary burst-search pipeline, while these two methods are used as auxiliary cross-verification tools rather than as independent parallel catalog-building pipelines. By comparing their candidate lists with the PRESTO detections on the same data set, we verified that there is no obvious population of missed bursts in the final sample. This process identified 17,356 bursts exceeding our detection threshold, defined by a fluence signal-to-noise ratio of SNR$_\mathrm{f}$ $\geq$ 12, where SNR$_\mathrm{f}$ denotes the signal-to-noise ratio of the burst fluence, i.e., the integrated burst flux over its duration, rather than the peak-flux signal-to-noise ratio (see \citealp{2025Zhangjs}). From this sample, we selected 6,384 bursts with S/N $\geq$20 for polarization analysis.

\subsection{Parameter Measurement}\label{subsec:ParamMears}

The time of arrival (TOA) for each burst was determined by measuring the time at which the cumulative fluence reached 50\% of its total value, following the burst identification criteria defined in \cite{2021Natur.LiD} and \cite{2025Zhanglx}, which classify multiple pulses as components of a single burst when a complex time-frequency structure is bridged by emission above 5$\sigma$. The final TOA is taken as the topocentric arrival time (MJD) at the 1.5 GHz reference. For the polarization analysis, the effective temporal limits of each burst are defined by its measured TOA and full width at half maximum (FWHM), adopting a time window of TOA$\pm 1.5\times$FWHM. The FWHM is obtained by fitting a Gaussian to the autocorrelation function of the same profile.

The dispersion measure (DM) of each burst was refined using the DM-power package \citep{2022Lin}. Unlike the traditional approach of maximizing the peak signal-to-noise ratio (S/N) of the frequency-summed profile, which can be biased by complex sub-burst drifting and frequency-dependent morphology, DM-power optimizes the DM in the Fourier domain. The method decomposes the dynamic spectrum into Fourier components along the time axis and measures the DM independently at different Doppler (Fourier) frequencies. Since short-timescale structures are more sensitive to residual dispersive delays, the algorithm effectively weights information across multiple timescales, yielding a more robust and high-precision DM measurement.

For bursts exhibiting ambiguous morphology in the dynamic spectrum or relatively low S/N, the multi-component optimization may become unstable and lead to overfitting. In such cases, we adopt the daily averaged DM values to ensure consistency across the sample. The derived DMs have a median value of 529.3 $\rm pc\ cm^{-3}$ with a standard deviation of 1.2 $\rm pc\ cm^{-3}$, consistent with the measurement reported by CHIME/FRB \citep{2025Shin}.

The burst frequency boundaries were determined using the CDF-based method proposed by \cite{2025ApJS.Xie}. For each burst, the spectrum was first constructed by integrating the on-pulse flux density over time for each frequency channel. The cumulative distribution function (CDF) of the spectrum was then calculated and smoothed using the asymmetrically reweighted penalized least squares (arPLS) algorithm to suppress noise fluctuations. The lower and upper frequency limits were defined as the frequencies where the first derivative of the normalized CDF rises above and subsequently falls below thresholds determined by the signal-to-noise ratio (S/N), specifically $1/(\mathrm{S/N})$ for the rising edge and $1–1/(\mathrm{S/N})$ for the falling edge. This procedure identifies the dominant emission band based on significant spectral power gradients rather than assuming a Gaussian spectral shape, and is therefore not equivalent to Gaussian fitting.

For bursts truncated by the observational band edges, a Gaussian extrapolation was applied to estimate the intrinsic bandwidth. For bursts with complex spectral morphologies or insufficient S/N for a reliable boundary identification, a boxcar bandwidth was adopted as a conservative estimate.

We applied the RM synthesis method to all 6,384 bursts with S/N$\geq$20, searching for rotation measures over the range from $-2000$ to $2000\ \mathrm{rad\ m^{-2}}$ \citep{2005A&A...441.1217B}. The adopted range was chosen as a conservative symmetric interval, comfortably wider than the RMs actually measured for FRB~20240114A. For the present FAST L-band channelization, intrachannel depolarization would become important only at much larger $|\mathrm{RM}|$, and therefore does not limit the current RM measurements. To ensure data reliability, we examined each burst for potential saturation effects in the FAST receiver. By inspecting the four polarization components (XX, YY, X$^*$Y, Y$^*$X) and identifying data points at the critical thresholds, we flagged and subsequently excluded bursts exhibiting genuine saturation signatures. This screening removed 253 bursts from the initial sample with S/N$\geq$20, leaving a final sample of 6,131 nonsaturated bursts with reliable RM measurements for the polarization analysis (see Appendix~\ref{sec:PolCal} for details). 

The observed RMs include a contribution from the Earth’s ionosphere, which adds a time-variable component to the intrinsic RM. Using the IonFR Python code and global ionospheric map (GIM) products \citep{2013ascl.soft03022S}, we calculated the ionospheric RM for each observing session and corrected the RM values (see Appendix \ref{sec:IonCorr} for details).

We additionally conducted a systematic search for Faraday conversion across all bursts. Faraday conversion is a propagation effect in a magnetized plasma with a perpendicular magnetic field, converting linear polarization into circular polarization.
Combined with Faraday rotation, if the Faraday conversion is observed, the detection would constrain plasma properties and magnetic field geometry along the line of sight \citep{2025ApJ...988..164W}. Our analysis found no significant evidence of conversion in any of the bursts in our sample, consistent with the results of \cite{2026arXiv260216409U}, indicating that the observed circular polarization is likely intrinsic to the emission mechanism rather than being generated by propagation effects.

\section{The Catalog Description}\label{sec:CataDesc}

The primary product of this work is a machine-readable catalog containing the measured parameters for the 6,131 bursts detected from FRB 20240114A. The catalog is structured as a table, with each row representing a single burst and each column a specific measured parameter, and the detailed description of the catalog columns is presented in Table \ref{tab:description}. An abridged version of the catalog is presented in Table \ref{tab:catalog_example} for illustration. The full catalog is available in online in CSV formats.

For the convenience of catalog users, we also provide estimates of the Galactic foreground contributions toward FRB~20240114A. Using the YMW16 and NE2025 Galactic electron-density models \citep{2017ApJ...835...29Y,2026arXiv260211838O}, we obtain Galactic interstellar dispersion measures of DM$_\mathrm{MW,ISM}$ = 38.82$\pm$4.78 pc~cm$^{-3}$ and 51.05$\pm$7.45 pc~cm$^{-3}$, respectively. The quoted uncertainties correspond to fractional model uncertainties of 12.32\% for YMW16 and 14.59\% for NE2025, estimated from pulsars with measured DMs at similar Galactic latitude by comparing the observed DMs with the corresponding model predictions. Using the YT2020 Galactic halo model \citep{2020ApJ...888..105Y}, we further estimate a halo contribution of DM$_\mathrm{MW,halo}$ = 45.05$^{+26.34}_{-16.62}$ pc cm$^{-3}$, where the uncertainty is obtained by applying the $\sim$0.2 dex (rms) fluctuation reported for the Milky Way halo DM in that model. We also estimate the Galactic Faraday rotation foreground from the Galactic Faraday rotation sky map \citep{2022A&A...657A..43H}, obtaining RM$_\mathrm{MW}$ = -14.55 $\pm$ 10.48 rad m$^{-2}$. These values are not burst-by-burst catalog entries, but source-level line-of-sight foreground references that may be useful for interpreting the measured DMs and RMs in the catalog.

\begin{deluxetable*}{rcl}
\tablewidth{0pt}
\tablecaption{Description of the Catalog Columns \label{tab:description}}
\tablehead{
\colhead{Column Name} & \colhead{Unit} & \colhead{Description}
}
\startdata
Burst ID & \dots & A unique identifier for each burst (which is consistent with \citealp{2025Zhangjs}).\\
MJD$_{\text{topo}}$ & $\text{day}$ & Topocentric arrival time (MJD) at the 1.5 GHz reference. \\
DM  & $\text{pc}\ \text{cm}^{-3}$ & Best-fit dispersion measure by DM-power. \\
W$_{\text{eff}}$ & ms & Full width at half maximum (FWHM) of the burst. \\
Bandwidth & MHz & The frequency bandwidth of the burst. \\
RM & $ \text{rad}\ \text{m}^{-2} $ & Rotation measure determined via RM synthesis. \\
$L/I$ & \% & Debiased linear polarization fraction. \\
$V/I$ & \% & Circular polarization fraction. \\
PA$_0$ & deg & The intrinsic polarization position angle at the infinite-frequency reference, corrected for RM. \\
S/N & \dots & Signal-to-noise ratio of the burst in the total intensity profile. \\
\enddata
\end{deluxetable*}

\begin{table}
	\centering
	\caption{\textbf{Sample table of burst parameters for FRB20240114A with S/N$\geq$20.}}
	\label{tab:catalog_example}
	\begin{tabular}{cccccccccc} 
		\\
		\hline
		Burst ID & MJD$_{\text{topo}}$ & DM  & RM  & $L/I$ & $V/I$ & W$_{\mathrm{eff}}$ & Bandwidth & PA$_0$ & S/N \\
         & & ($\text{pc}/\text{cm}^3$) & ($\text{rad}/\text{m}^2$) & (\%) & (\%) & ($\text{ms}$) & ($\text{MHz}$) & (deg) & \\
		\hline

        B00001 & 60337.23827975 & 529.5$\pm$0.1 & 328.5$^{+1.6}_{-1.6}$ & 96.3$\pm$2.1 & -7.1$\pm$1.5 & 3.51$\pm$0.23 & 188.3$\pm$10.6 & -7.8 & 29.7\\
        B00007 & 60337.25012753 & 527.7$\pm$0.1 & 317.5$^{+1.7}_{-1.7}$ & 97.3$\pm$2.8 & -12.3$\pm$2.0 & 3.02$\pm$0.21 & 207.7$\pm$6.2 & 46.3 & 38.3\\
        B00009 & 60338.21737789 & 528.5$\pm$0.1 & 326.5$^{+2.8}_{-2.9}$ & 97.1$\pm$3.1 & -26.1$\pm$2.3 & 1.62$\pm$0.06 & 319.2$\pm$8.1 & 19.0 & 46.9\\
        B00011 & 60338.22518088 & 527.2$\pm$0.6 & 335.5$^{+6.4}_{-5.3}$ & 60.4$\pm$3.8 & -82.4$\pm$4.2 & 2.52$\pm$0.14 & 74.5$\pm$6.7 & 7.3 & 20.4\\
        B00012 & 60338.22802442 & 528.3$\pm$0.1 & 330.5$^{+2.2}_{-2.1}$ & 98.8$\pm$2.1 & 1.0$\pm$1.5 & 2.0$\pm$0.07 & 158.7$\pm$21.2 & 73.1 & 42.8\\
        B00018 & 60338.23511480 & 531.1$\pm$0.6 & 330.5$^{+1.9}_{-1.9}$ & 102.3$\pm$2.8 & -11.6$\pm$1.9 & 3.96$\pm$0.2 & 155.5$\pm$4.0 & 11.8 & 22.0\\
        B00019 & 60338.23556485 & 529.1$\pm$0.1 & 327.5$^{+2.2}_{-2.3}$ & 92.1$\pm$2.6 & -0.7$\pm$1.9 & 2.47$\pm$0.1 & 174.1$\pm$21.5 & 8.7 & 33.0\\
        B00028 & 60341.19895450 & 530.4$\pm$0.5 & 343.5$^{+2.4}_{-2.3}$ & 54.6$\pm$2.5 & 11.2$\pm$2.2 & 3.65$\pm$0.2 & 121.3$\pm$6.6 & -53.4 & 20.5\\
        B00031 & 60349.15748811 & 529.0$\pm$0.3 & 348.5$^{+2.6}_{-2.7}$ & 98.4$\pm$1.8 & 18.4$\pm$1.3 & 1.94$\pm$0.09 & 90.2$\pm$14.3 & -48.8 & 28.8\\
        B00032 & 60349.15783813 & 528.5$\pm$0.1 & 350.5$^{+0.7}_{-0.7}$ & 96.1$\pm$1.2 & -6.5$\pm$0.9 & 4.91$\pm$0.16 & 237.6$\pm$9.4 & 23.6 & 34.6\\
        \multicolumn{9}{c}{\vdots} \\
        B17290 & 60824.89193507 & 530.9$\pm$0.1 & 246.1$^{+1.0}_{-1.1}$ & 99.3$\pm$2.2 & -5.1$\pm$1.5 & 3.09$\pm$0.02 & 276.4$\pm$5.9 & 7.4 & 56.0\\
        B17310 & 60824.89230168 & 530.0$\pm$0.1 & 231.9$^{+7.0}_{-5.1}$ & 86.3$\pm$9.4 & 1.8$\pm$7.1 & 2.81$\pm$0.03 & 325.1$\pm$12.6 & 52.6 & 57.4\\
        B17313 & 60824.89238065 & 529.8$\pm$0.1 & 250.7$^{+1.3}_{-1.3}$ & 99.2$\pm$2.2 & -10.8$\pm$1.6 & 1.98$\pm$0.01 & 458.9$\pm$9.4 & -50.2 & 152.0\\
        B17315 & 60824.89327416 & 529.6$\pm$0.2 & 248.1$^{+1.4}_{-1.7}$ & 103.8$\pm$3.0 & 1.4$\pm$2.1 & 3.99$\pm$0.02 & 442.5$\pm$14.7 & -11.2 & 54.7\\
        B17317 & 60824.89356171 & 530.2$\pm$0.1 & 242.7$^{+1.0}_{-0.9}$ & 91.5$\pm$1.9 & -22.9$\pm$1.5 & 3.04$\pm$0.03 & 204.1$\pm$10.3 & 35.1 & 53.6\\
        B17318 & 60824.89386021 & 529.8$\pm$0.1 & 245.3$^{+0.4}_{-0.4}$ & 97.7$\pm$0.7 & 9.9$\pm$0.5 & 3.65$\pm$0.01 & 255.5$\pm$8.8 & 59.2 & 177.9\\
        B17328 & 60824.89416654 & 529.4$\pm$0.2 & 243.5$^{+0.7}_{-0.7}$ & 100.2$\pm$1.5 & -2.9$\pm$1.0 & 2.55$\pm$0.02 & 195.6$\pm$6.7 & -64.3 & 106.0\\
        B17332 & 60824.89625329 & 530.6$\pm$0.1 & 240.1$^{+4.5}_{-4.0}$ & 90.9$\pm$8.7 & 16.1$\pm$6.5 & 2.19$\pm$0.07 & 287.8$\pm$4.2 & 11.0 & 26.4\\
        B17344 & 60824.89625000 & 530.5$\pm$0.1 & 239.5$^{+1.4}_{-1.3}$ & 94.4$\pm$4.7 & 0.9$\pm$3.4 & 2.12$\pm$0.05 & 213.0$\pm$8.6 & 33.1 & 40.8\\
        
		\hline
	\end{tabular}
    \vspace{4mm}
	\begin{minipage}{1\linewidth}
		\footnotesize
		\textit{Note.} All uncertainties in this table are rounded to one decimal place, following \cite{2025Zhangjs} for consistency with previously published results. For a small subset of bursts whose DM uncertainties would be rounded to 0.0 under this convention, we uniformly adopt a conservative floor value of 0.1 pc~cm$^{-3}$to avoid reporting unphysical zero-valued uncertainties. The quoted RM uncertainties are empirical estimates based on the RM-synthesis peak width and burst S/N. An additional systematic uncertainty comparable to one RM-search grid step ($\sim$ 0.2 rad~m$^{-2}$) may also be considered.
	\end{minipage}
\end{table}

\section{Data Quality and Global Properties}\label{sec:quality}

To validate the data quality and present an overview of the source's behavior, we show the distributions of and correlations between key parameters from the catalog. This analysis serves to characterize the dataset for future scientific work and does not attempt a detailed physical interpretation.

\subsection{Parameter Distributions}\label{subsec:ParaDist}

\begin{figure*}[ht!]
\plotone{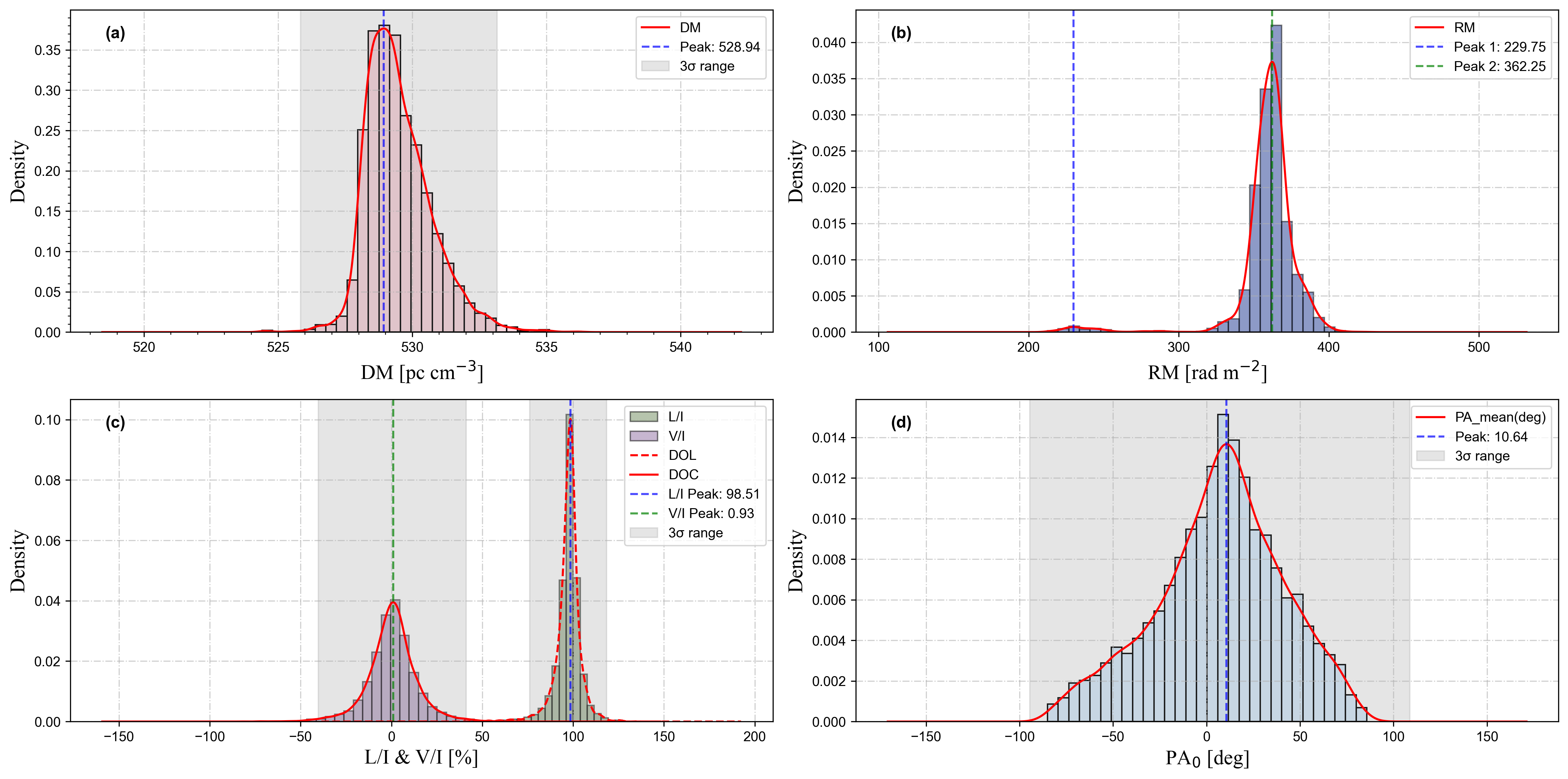}
\caption{\textbf{Distributions of key parameters for the 6,131 bursts from FRB 20240114A.} (a) DM. The distribution is narrow and peaks near 528.9 $\rm pc\ cm^{-3}$. (b) RM. The distribution shows a complex, non-Gaussian shape with a primary peak near 362.2 $\rm rad\ m^{-2}$ and a secondary peak near 229.8 $\rm rad\ m^{-2}$. (c) Linear ($L/I$, green hist and dashed line) and circular ($V/I$, purple hist and solid line) polarization fractions. (d) Intrinsic polarization position angle (PA$_0$), corrected for Faraday rotation. \label{fig:dist}}
\end{figure*}

Figure \ref{fig:dist} presents the histograms of the primary measured parameters for the entire burst sample.

\begin{itemize}
    \item \textbf{Dispersion Measure (DM, upper left panel)}: The DM distribution peaks at approximately 528.9 $\rm pc\ cm^{-3}$, consistent with previously reported values for this source \citep{2025Shin}. The narrow spread of the distribution demonstrates the stability of the line-of-sight integrated electron density over the 16-month observing campaign.
    \item \textbf{Rotation Measure (RM, upper right panel)}: The RM distribution exhibits a complex, non-Gaussian structure. The primary peak is located near 362.2 $\rm rad\ m^{-2}$, corresponding to the initial stable phase of the source. A secondary, smaller but visually distinct peak is present at a lower RM of approximately 229.8 $\rm rad\ m^{-2}$. This bimodality directly reflects the significant temporal evolution of the RM (see Fig. \ref{fig:overview}a), where the secondary peak corresponds to bursts detected during the later phase of rapid RM decline and a subsequent clustering of activity.
    \item \textbf{Linear/Circular Polarization Fraction ($L/I$, $V/I$, lower left panel)}: The linear polarization fractions are systematically high, with a 3$\sigma$ lower bound of 76\%, a behavior commonly seen among repeating FRBs and suggestive of a magnetically ordered emission region. The distribution of the absolute circular polarization fraction is peaked at low values, with a majority of bursts below 40\%. A tail of bursts extends to higher values, suggesting diverse emission properties or propagation effects within the sample.
    \item \textbf{Intrinsic Polarization Angle (PA$_0$, lower right panel)}: The distribution of the intrinsic polarization position angles is broad, spanning the full $-90^\circ- 90^\circ$ range with a peak approximately around 10$^\circ$. It is not uniform, but exhibits a noticeable, non-Gaussian excess. While a concentration of bursts suggests a possible preferential orientation in the emission or projection geometry during parts of the observing campaign, the broad underlying distribution indicates that this preference is not stable or exclusive. The overall pattern is inconsistent with a single, stable global magnetic field direction governing all bursts.
\end{itemize}

\subsection{Parameter Correlations}\label{subsec:ParaCorr}

\begin{figure*}[ht!]
\plotone{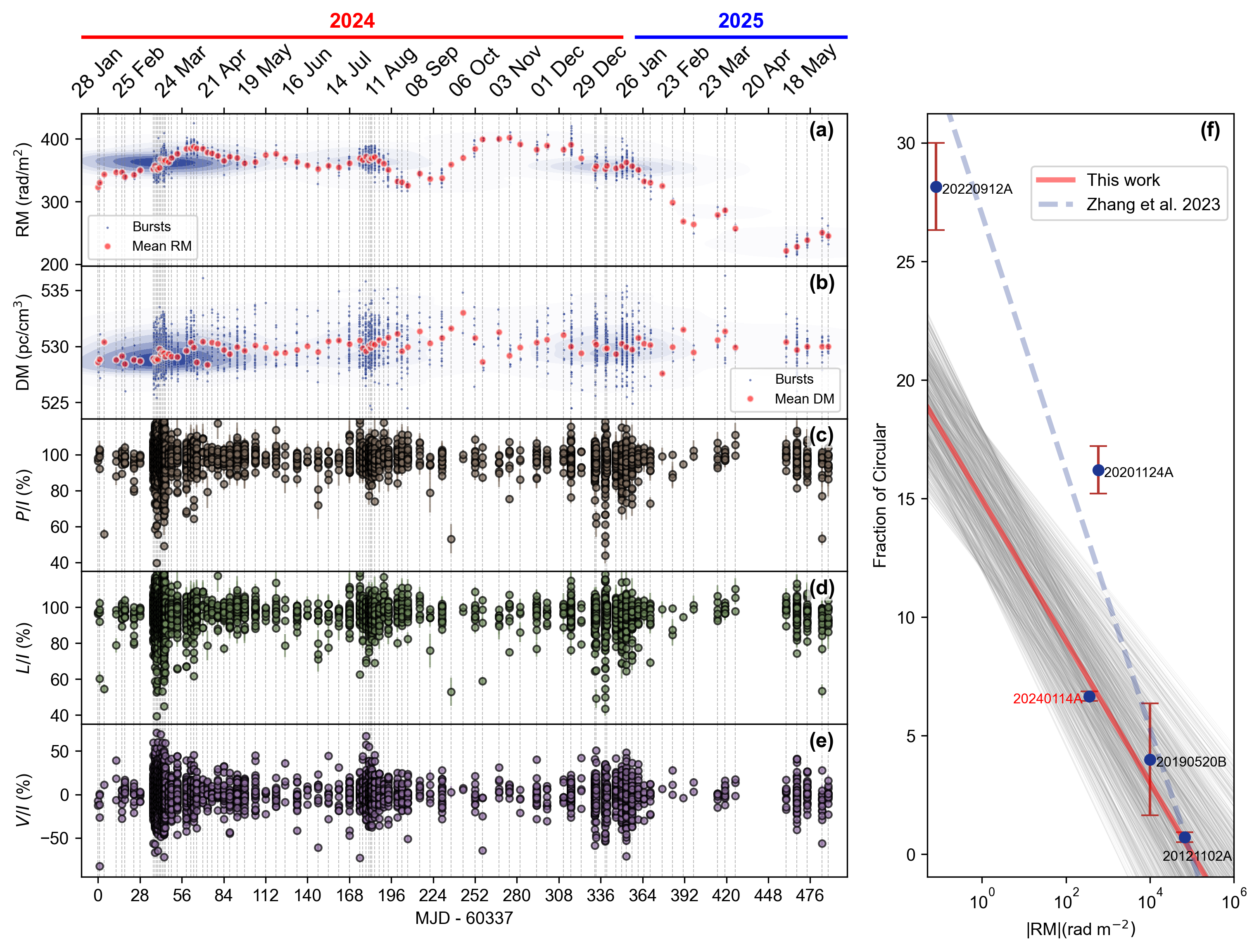}
\caption{\textbf{Temporal evolution of burst properties.} Panels (a) to (e) share a common horizontal axis (Topocentric MJD). (a) RM as a function of time. The blue contour is the 2D kernel density estimation (KDE) of the RM. (b) DM over time. The blue contour is the 2D KDE of the DM. (c) Total polarization fraction over time. (d) Linear polarization fraction over time. (e) Circular polarization fraction ($V/I$) over time. Panel (f) (right) shows the relationship between the absolute circular polarization fraction ($|\textrm{V}|$/I) and the absolute value of the RM ($|\textrm{RM}|$). The red solid line is the fitting for 5 repeating FRBs including FRB 20240114A, while the blue dashed line is the fitting for 4 repeating FRBs not including FRB20240114A. Plotted in gray are the results of the allowable values from the Markov Chain Monte Carlo (MCMC) run. \label{fig:overview}}
\end{figure*}

\begin{figure*}[ht!]
\plotone{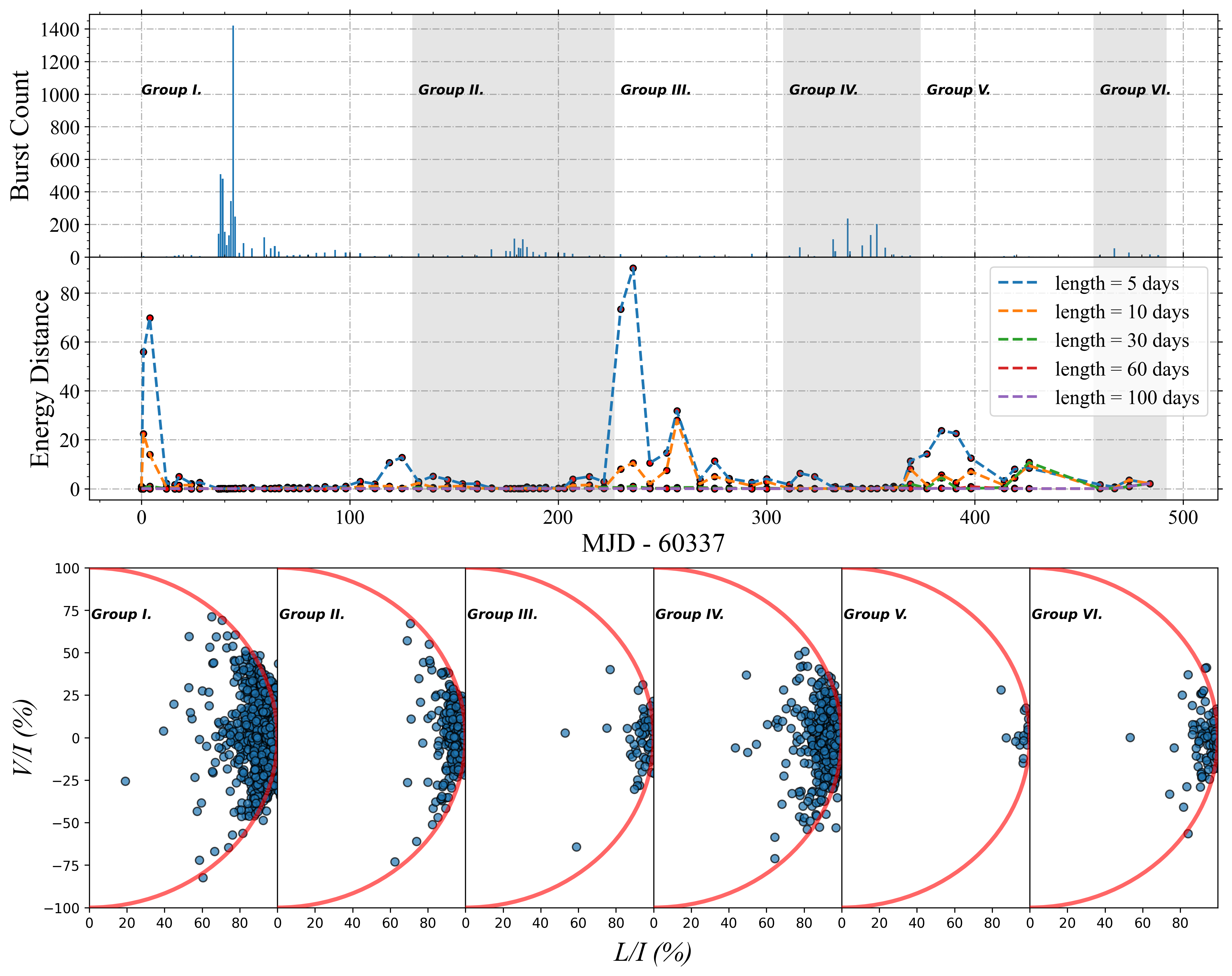}
\caption{\textbf{Temporal evolution of linear and circular polarization fraction distributions.} (Top) Histogram of the number of bursts with signal-to-noise ratio $\geq$ 20 per observing session. The timeline is divided into six alternating shaded/unshaded sessions (I-VI) based on burst activity. (Middle) Energy Distance (a statistical measure of similarity between distributions) calculated between the polarization fraction distributions of cumulative bursts in consecutive observing sessions with different time-scale cumulative windows. Some peaks in energy distance correspond to sessions with a sudden, large change in the number of data points (bursts) rather than a genuine shift in the underlying distribution shape. (Bottom) Temporal evolution of the polarization fraction distributions visualized for sessions I-VI using a semicircle diagram. \label{fig:DoPevo}}
\end{figure*}

\begin{figure*}[ht!]
\plotone{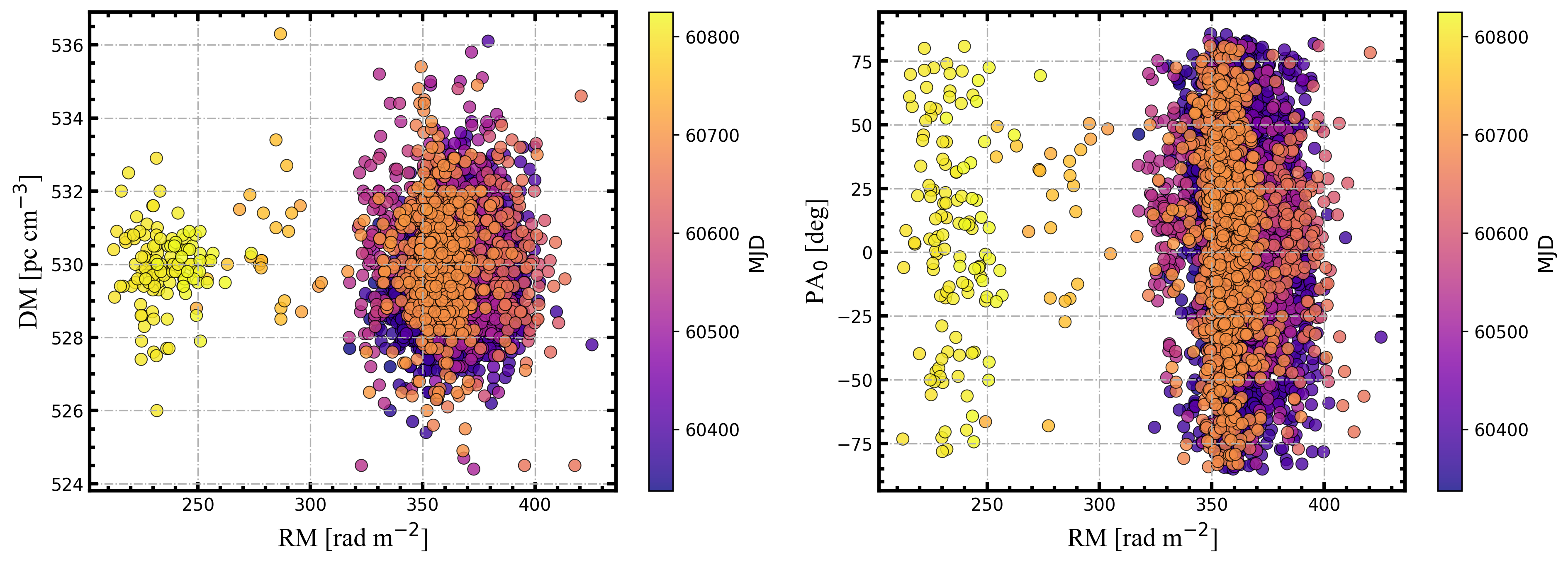}
\caption{\textbf{Rotation measure plotted versus dispersion measure and polarization position angle.} (Left) RM vs. DM, with points color-coded by topocentric MJD.  (Right) RM vs. intrinsic polarization position angle (PA$_0$), similarly color-coded.\label{fig:DM-RM}}
\end{figure*}

\begin{figure*}[ht!]
\plotone{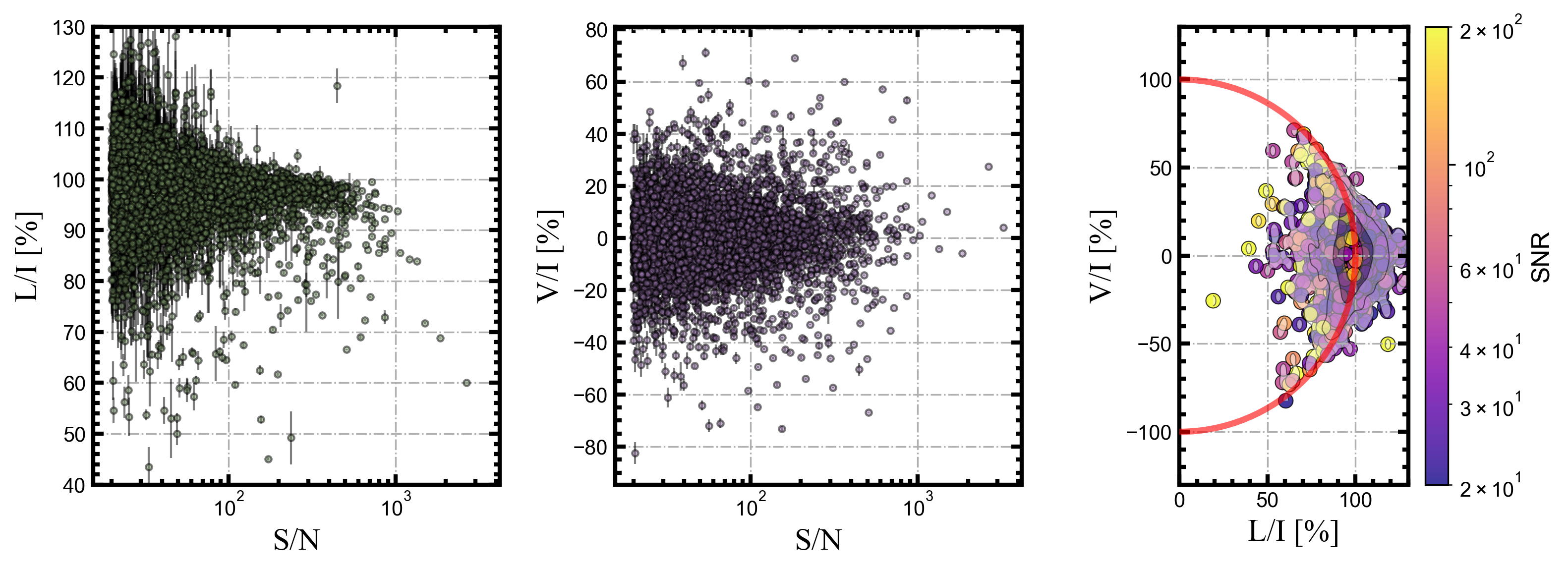}
\caption{\textbf{Relationships involving polarization fractions and signal-to-noise ratio (S/N).} (a) Linear polarization fraction ($L/I$) vs. S/N. (b) Circular polarization fraction ($V/I$) vs. S/N. (c) Relationship between linear and circular polarization fractions. Points are color-coded by S/N, and number-density contours (gray lines) illustrate the distribution of the large dataset. The solid red semi-circle represents the theoretical physical limit for 100\% polarized emission. \label{fig:DOP-SNR}}
\end{figure*}

Figures \ref{fig:overview}-\ref{fig:DOP-SNR} explore the relationships among several key parameters to assess the overall consistency of the measurements and to identify global trends in the burst sample.

\begin{itemize}
    \item \textbf{RM evolution (Figure \ref{fig:overview})}: Among the parameters shown in Figure \ref{fig:overview}, RM is the only one that exhibits a pronounced long-term evolution, consistent with the trend previously noted by \cite{2026arXiv260216409U}. As shown in panel (a), RM as a function of topocentric arrival time displays an initial period of relative stability followed by a significant linear decay. In contrast, the corresponding time evolutions of DM, total polarization fraction ($P/I$), linear polarization fraction ($L/I$), and circular polarization fraction ($V/I$) shown in panels (b)–(e) do not display a similarly distinct transition, and instead remain comparatively stable or scattered without a clear monotonic trend. This RM evolution will be the focus of a more detailed analysis in a subsequent paper.
    \item \textbf{Polarization Fraction Distribution evolution (Figure \ref{fig:DoPevo})}: The top panel shows the observed burst number (with S/N $\geq$ 20) as a function of time, indicating significant variations in activity level. The middle panel quantifies the similarity between the two-dimensional (linear vs. circular) polarization fraction distributions of cumulative bursts in consecutive observing sessions with different time-scales using the Energy Distance metric (see Appendix \ref{sec:ED}). A larger Energy Distance corresponds to lower similarity between the two distributions. The notable peaks in this metric are found to correlate with abrupt, large changes in the number of detected bursts (as seen in the top panel), rather than signifying a fundamental change in the shape of the underlying polarization distribution itself. This conclusion is visually and statistically confirmed by the semicircle diagram in the bottom panel. Across the entire $\sim$16-month span, the dense locus of points consistently resides near the high-linear-polarization, low-circular-polarization end of the theoretical limit (the semicircle boundary). The core shape and concentration of this two-dimensional distribution show no systematic evolution correlated with the drastic RM change or the variable burst rate.
    
    \item \textbf{RM versus DM (Figure \ref{fig:DM-RM} left panel)}: No strong correlation is observed between RM and DM across the entire dataset (Pearson correlation coefficient r=-0.023). This lack of correlation suggests that the RM variability is not controlled by changes in the total electron column density alone. Instead, it is consistent with a scenario in which the Faraday rotation is dominated by variations in the line-of-sight magnetic field component, including possible changes in its strength or orientation, within a complex magneto-ionic environment.
    \item \textbf{RM versus PA$_0$ (Figure \ref{fig:DM-RM} right panel)}: Similar to DM, intrinsic polarization position angle (PA$_0$) is also separable into two groups chronologically and exhibits no strong correlation (Pearson correlation coefficient r=-0.049). The distribution in PA$_0$ is broader for a given RM, resulting in two vertical columns rather than compact clusters. This indicates that while the RM (probing the line-of-sight magnetic field) changed systematically, the intrinsic PA (related to the emitting region magnetic geometry) retained a wide range of orientations throughout the evolution.
    \item \textbf{Polarization Fraction versus S/N (Figure \ref{fig:DOP-SNR})}: The linear and circular polarization fractions are plotted against the burst S/N (left and middle panel). As expected, the measurement uncertainty increases at low S/N. The absence of any spurious trends at high S/N confirms the robustness of our polarization calibration and debiasing procedures (see Appendix \ref{sec:RMsyn}). The right panel shows the relationship between the linear ($L/I$) and circular ($V/I$) polarization fractions. A theoretical semi-circle with a radius of 100\%, centered on the origin and extending in the positive $L/I$ direction, is overlaid. This curve represents the physical limit for fully polarized emission. The points are color-coded by the burst S/N, and number-density contours are added to illustrate the distribution of the large dataset. Two key features are evident: (1) The vast majority of data points, especially those with high number density, cluster near the intersection of the semi-circle and the $L/I$ axis (i.e., high linear and low circular polarization). (2) Points lying outside the theoretical semi-circle are predominantly those with low S/N. This is fully consistent with expectations, as measurement noise can spuriously push low-S/N points beyond the physical boundary, further validating our error estimation and the overall quality of the polarimetric measurements.    
\end{itemize}

In addition to the above parameters, we also calculated the proportion of bursts exhibiting significant circular polarization (degree of circular polarization, $|V|/I$ $\geq$ 10\%) in our sample. For FRB 20240114A, 1,157 out of 17,356 detected bursts met this criterion, yielding a proportion of 6.67\% (see Figure \ref{fig:overview} panel f). This measurement was examined within the context of the tentative anti-correlation between the significant circular polarization fraction and the absolute value of the source’s rotation measure ($|\textrm{RM}|$) suggested by \cite{2023ApJ.ZhangYK}. A power-law fit to our updated sample yields an index of $-2.98 \pm 0.80$. With this fit, the associated uncertainty envelope encompasses the data points for FRB 20190520B and FRB 20121102A \citep{2023ApJ.ZhangYK, 2022SciBu..67.2398F}, while those for FRB 20220912A and FRB 20201124A \citep{2023ApJ.ZhangYK} lie significantly outside of it. Should an underlying physical correlation exist, the fact that only a subset of sources aligns with the best-fit trend may point to intrinsic diversity within the repeating FRB population. Sources following the relation might originate in environments where the observed circular polarization is dominantly shaped by the same large-scale, ordered magneto-ionic medium responsible for the Faraday rotation \citep{2022NaturNiuCH,2023Sci.Anna-Thomas}. In contrast, the significant deviation of other sources could imply that their polarization properties are governed by additional or distinct physical processes. For instance, complex local Faraday screens, burst emission geometries that favor strong intrinsic circular polarization, or magnetospheric propagation effects could decouple the observed circular polarization from the bulk RM in these objects \citep{2021ApJ.Hilmarsson,2022Natur.XuH,2023ApJMckinven}. This dichotomy underscores that any universal relationship between circular polarization and RM is likely nuanced and modulated by source-specific conditions. It should be noted that this alignment with the trend for FRB 20240114A does not necessarily imply that the generation mechanism of circular polarization is identical to that of sources such as FRB 20121102A. Therefore, the circular polarization of FRB 20240114A may still be attributed primarily to the intrinsic emission mechanism itself, rather than being dominated by propagation effects.

In summary, the power-law fit reveals a shallower index for the $|V|/I$–$|\textrm{RM}|$ relation, which is consistent with only a subset of repeating FRBs. This suggests that any such correlation is not universal but is likely modulated by source-specific local environments, as evidenced by the diverse RM behaviors predominantly attributed to proximate magneto-ionic screens. The dichotomy between sources that follow and those that deviate from the trend points to intrinsic diversity in either the emission mechanism or the magneto-ionic properties of the immediate vicinities of repeaters \citep{2023MNRAS.YangYP,2024MNRAS.Uttarkar}. Continued observations of FRB~20240114A, together with comparisons across a larger sample of repeating FRB sources, will be crucial for disentangling these physical effects.

\begin{figure}
    \centering
    \includegraphics[width=0.9\textwidth]{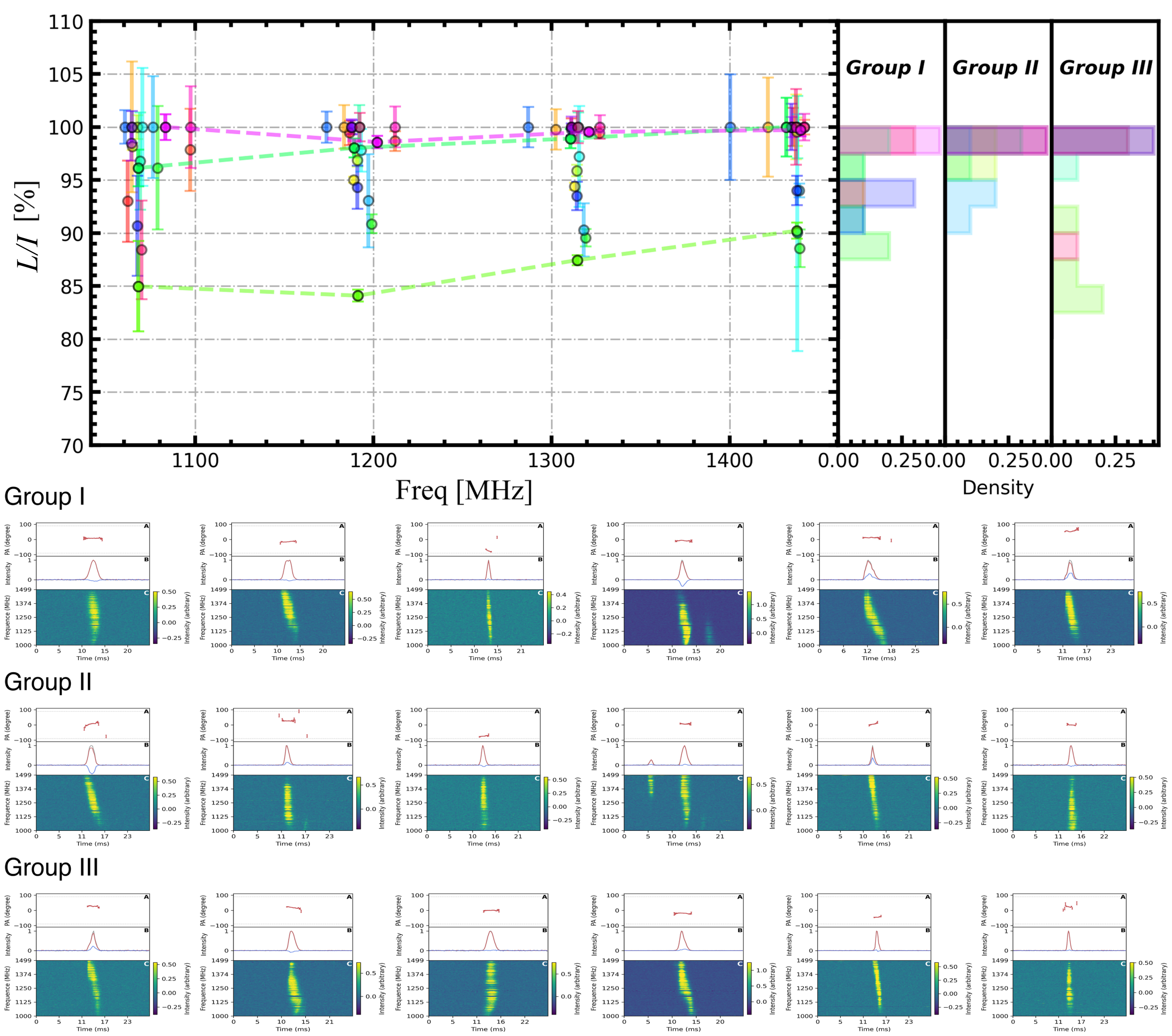}
    \caption{\textbf{Test for frequency-dependent depolarization ($\sigma_{\text{RM}}$)}. For 18 bright, broadband bursts with high signal-to-noise ratio, we subdivided each burst’s bandwidth into segments, independently measured the RM and linear polarization fraction ($L/I$) in each segment, and plotted $L/I$ against the center frequency of the segment. Points are color-coded by burst ID. To illustrate the frequency dependence of the $L/I$, three bursts were randomly selected and their points were plotted as line plots. To avoid overlap, the 18 bursts are randomly divided into three groups for display. The 18 plots in the bottom panel show the dynamic spectrum and polarization profile for these bursts.}
    \label{fig:sigmaRM}
\end{figure}

Having characterized the parameter distributions and correlations for FRB~20240114A, we next examine the frequency-dependent depolarization to explore the burst’s magneto-ionic environment. A random magnetic field component within the Faraday rotating screen would cause frequency-dependent depolarization, characterized by a decrease in the observed linear polarization fraction ($L/I$) towards lower frequencies. In the classical \cite{1966MNRAS.133...67B} framework for depolarization by Faraday dispersion, the fractional reduction in linear polarization can be written as
\begin{equation}
    f_{\text{RM}\ scattering} \equiv 1\ -\exp(-2\lambda^4\sigma^2_{\text{RM}}),
\end{equation}
where $f_{\text{RM}\ scattering}$ is the fractional reduction in the linear polarization amplitude, $\sigma_{\text{RM}}$ is the standard deviation of RM and $\lambda$ is the wavelength. This formalism has since been applied to FRBs by \cite{2022Sci.FengY}, \cite{2026MNRAS.545f1997U}, and \cite{2026arXiv260222309P}. To test for this effect, we selected 18 bright bursts with high signal-to-noise ratio and broad bandwidth. The dynamic spectra and polarization profiles of these 18 bursts are shown in Figure \ref{fig:sigmaRM}. Each bursts’ band was subdivided into several frequency sub-bands. We independently performed RM synthesis and measured $L/I$ within each sub-band. As shown in Figure \ref{fig:sigmaRM}, the derived $L/I$ values for all sub-bands across all selected bursts cluster near 100\%, with no systematic trend as a function of frequency. To better illustrate whether the $L/I$ varies significantly with frequency, we plotted the $L/I$ distribution of slices for each burst (left panel of Figure \ref{fig:sigmaRM}). To avoid overlap, the bursts were randomly divided into three groups. If the $L/I$ distribution of a burst exhibits significant broadening, it would indicate a pronounced frequency evolution of the $L/I$, whereas a concentrated distribution implies that the $L/I$ is nearly independent of frequency. The results show that the $L/I$ distributions of all 18 bursts are relatively concentrated, indicating that the $L/I$ remains nearly constant across frequency. This absence of significant depolarization allows us to place an upper limit on any random magnetic field component (often quantified as $\sigma_{\text{RM}}$) within the main Faraday screen. Based on our Bayesian analysis of 18 bursts, we derive a 95\% upper limit of $\sigma_\mathrm{RM}$ $<$ 0.51 rad m$^{-2}$.This suggests that the main Faraday screen is largely ordered and exhibits minimal internal Faraday dispersion, at least for the selected bursts analysed here. A broader population-level test based on burst central frequency, bandwidth, and $L/I$ would be valuable, but is not yet sufficiently robust for the present data set because our current CDF-based bandwidth measurements do not always yield reliable central-frequency estimates, particularly for bursts truncated by the band edges. In addition, for multi-component bursts, differences in $L/I$ among distinct temporal or spectral components may contaminate a depolarization search by introducing apparent frequency-dependent behavior that does not arise from a single, uniform Faraday screen.

We additionally searched for signatures of Faraday conversion in bursts with sufficient S/N and frequency coverage. Specifically, we examined whether the circular polarization fraction exhibits any systematic frequency- or wavelength-dependent evolution within individual bursts. We do not find any clear or definitive evidence for such behavior in the present sample. This is also consistent with the result of \cite{2026arXiv260216409U}.

\section{Discussion}\label{sec:discussion}

\subsection{The Origin of RM Evolution}\label{subsec:origin}

\begin{figure}
    \centering
    \includegraphics[width=0.9\textwidth]{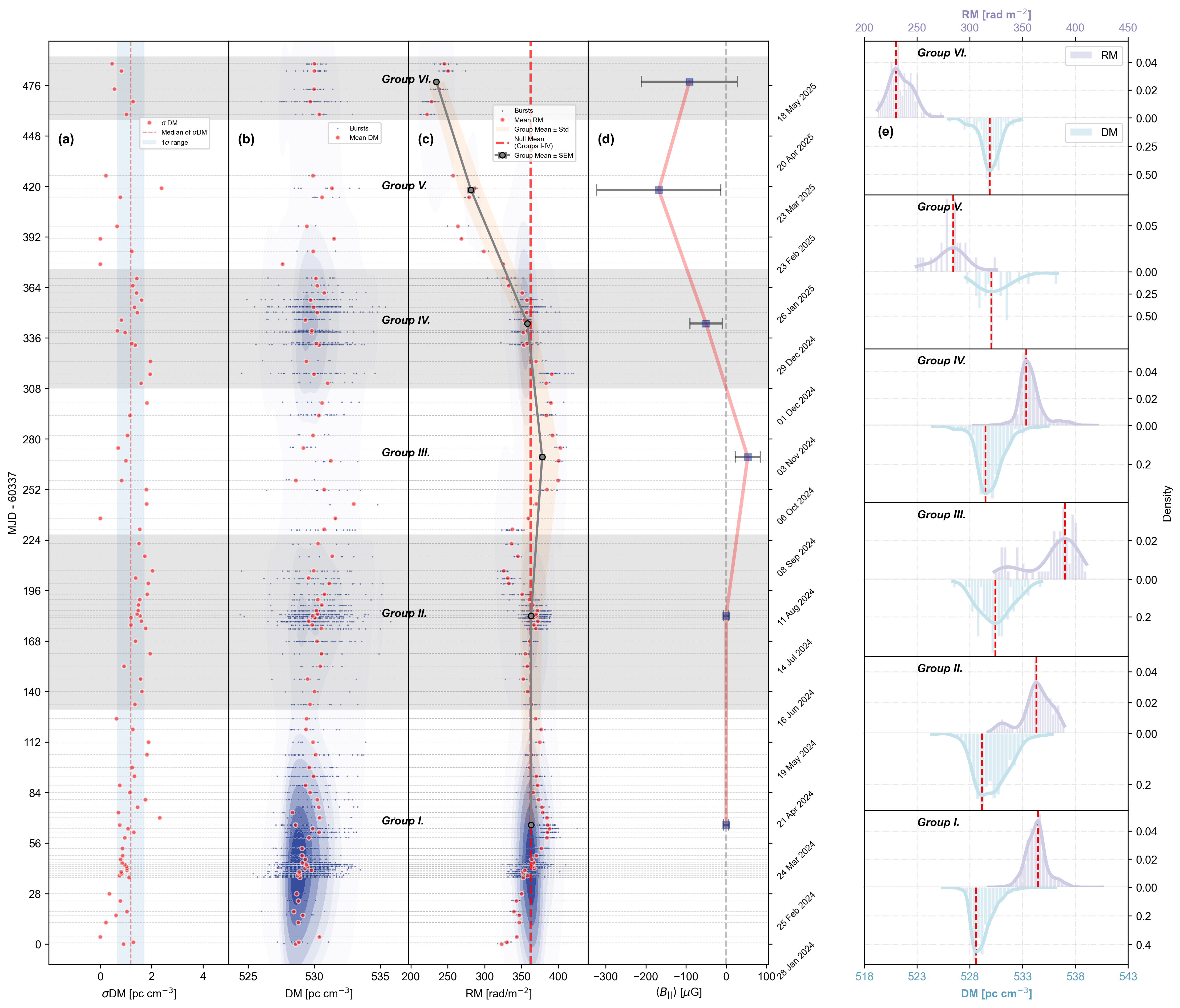}
    \caption{\textbf{Temporal evolution of DM, RM and $\langle B_{\parallel}\rangle$.} Panel (a): The temporal evolution of $\sigma_{\rm DM}$, where $\sigma_{\rm DM}$ denotes the standard deviation of the burst DM measurements within each observing interval. The red dashed line represents the median, and the light blue shaded area indicates the 1-$\sigma$ range. Panel (b) and (c): The temporal evolution of DM and RM is displayed with MJD on the vertical axis. The blue contours represent the two-dimensional kernel density estimation (KDE) of the data points. The timeline is divided into six alternating shaded/unshaded sessions (I-VI) based on burst activity. In the RM panel, gray solid lines with error bars show the Standard Error of the Mean (SEM) for bursts within contiguous time bins, and the orange band indicates the standard deviation. The red dashed line marks the mean RM under the null hypothesis of no magnetic field evolution. Panel (d): The evolution of $\langle B_{\parallel}\rangle$ for six sessions. Panel (e): The corresponding distributions of RM (top sub-panel) and DM (bottom sub-panel) for each of the six sessions (I-VI, from bottom to top). The red dashed lines indicate the peak of each distribution.}
    \label{fig:DM-RMevo}
\end{figure}

Our session-based analysis (Figure \ref{fig:DM-RMevo}) starkly contrasts the temporal behaviors of DM and RM. The DM remains stable across all six activity-defined sessions, with its distribution peak and width showing negligible changes (Figure \ref{fig:DM-RMevo}, panel e). This indicates that no strong secular DM evolution is detected across the six activity-defined sessions. To quantify the additional burst-to-burst DM scatter beyond the formal fitting uncertainties, we compared the burst-level DMs with the corresponding daily-average reference DM values and wrote the observed daily scatter as $\sigma_{\rm obs,daily}^2=\sigma_{\rm measure}^2+\sigma_{\rm sys}^2+\sigma_{\rm int}^2$, where $\sigma_{\rm measure}$ is the formal measurement contribution, $\sigma_{\rm sys}$ represents the morphology-related excess variance, and $\sigma_{\rm int}$ denotes any intrinsic day-scale DM variation. For our sample, we obtain $\sigma_{\rm obs,daily}=1.10~{\rm pc~cm^{-3}}$ and $\sigma_{\rm measure}=0.84~{\rm pc~cm^{-3}}$, implying $(\sigma_{\rm sys}^2+\sigma_{\rm int}^2)^{1/2}\approx0.70~{\rm pc~cm^{-3}}$. Under the practical assumption that $\sigma_{\rm int}=0$, this yields an upper limit of $\sigma_{\rm sys}\lesssim0.70~{\rm pc~cm^{-3}}$. We therefore interpret the absence of an obvious long-term DM trend as a non-detection above the burst-level variance floor, rather than as evidence for perfectly constant DM. This interpretation is also supported by the recent FAST study of \cite{2026arXiv260401825F}, who argued that part of the apparent burst-to-burst DM scatter in repeating FRBs can be morphology-induced (``pseudo DM") rather than reflecting true propagation changes. Our result for FRB~20240114A is qualitatively consistent with this picture.  In dramatic contrast, the RM undergoes a highly significant, monotonic decay of approximately 200 $\rm rad\ m^{-2}$. The shift in the RM distribution peak between sessions is visually clear (Figure \ref{fig:DM-RMevo}, panel e).

To quantify the significance of this decline, we performed a Monte Carlo test using the burst-level RM measurements from sessions I–IV as the null hypothesis of a stable RM distribution. Specifically, we combined all RM values from these four sessions and used their sample mean and standard deviation to define the parent distribution. We then generated 10,000 mock samples, each containing the same number of bursts as observed in sessions V–VI, by drawing random values from this null distribution. For each mock sample, we calculated its mean RM and compared the resulting distribution of simulated means with the observed mean RM of sessions V–VI. The observed value lies far below the simulated distribution, and none of the 10,000 realizations reproduces such a low mean, implying p $<$ 10$^{-4}$ under the null hypothesis of no RM evolution.

The evolution of the line-of-sight magnetic field component $\langle B_{\parallel}\rangle$ can be quantitatively estimated from the changes in RM and DM, which is given by $\langle B_{\parallel}\rangle \simeq 1.2 \times \Delta \text{RM}/\Delta \text{DM}\ \rm \mu G $, where $ \Delta \text{RM} $ and $ \Delta \text{DM} $ are the differences in RM and DM between different sessions. Applying this to the significant RM decay observed between activity sessions, we find that the nominal values of $\langle B_{\parallel}\rangle$ vary in a manner qualitatively consistent with the RM evolution (Figure \ref{fig:DM-RMevo}, panel d). However, the uncertainties are large, and the current data do not provide statistically significant evidence for a departure from a constant temporal evolution of $\langle B_{\parallel}\rangle$. The uncertainties of $\langle B_{\parallel}\rangle$ are derived from error propagation of the measurement errors in RM and DM. The stable standard deviation of DM over time (Figure \ref{fig:DM-RMevo}, panel a) argues against the injection of new, clumpy plasma, supporting the interpretation that the changing RM and $\langle B_{\parallel}\rangle$ stem from the evolution of an existing magneto-ionic environment.

This inferred local line-of-sight magnetic field can also be compared with similar estimates discussed for other repeating FRBs. For FRB~20240114A, our session-based estimates span from values consistent with zero at early times to tens and up to $\sim10^2\ \mu \mathrm{G}$ at later epochs, with nominal values ranging from about -168 to +54 $\mu$G (i.e., $\sim$0.05-0.17 mG in magnitude). This places FRB~20240114A below the more strongly magnetized mG-class environments inferred for FRB~20121102A \citep{2018Natur.Michilli}, FRB~20190520B \citep{2023SciAnna-Thomas}, FRB~20190417A \citep{2026ApJ...996L..16M}, and FRB~20220529A \citep{2026arXiv260222309P}, and also somewhat below FRB~20240619D \citep{2026MNRAS.546ag090O}. In this sense, FRB~20240114A appears less extreme than highly magnetized systems, while still indicating a dynamically evolving magneto-ionic environment. More broadly, it is comparable to the class of repeaters for which RM and DM evolution imply a local environment that is clearly magnetized but not as extreme as the most strongly Faraday-active sources.

The coherent, large-scale decay in both RM and the inferred $\langle B_{\parallel}\rangle$ points to a global evolution of the magneto-ionic environment surrounding the FRB source. This could be due to the expansion of a magnetized shell (e.g., a supernova remnant or a pulsar wind nebula \citep{2018Natur.Michilli}), the traversal of the source through a gradient in a large-scale magnetic field, or the secular evolution of a magnetized stellar wind from a binary companion \citep{2022Natur.XuH}.

A useful phenomenological comparison is FRB~20180916B, whose RM evolution also exhibits an initially relatively stable phase followed by a more secular change \citep{2023ApJ...950...12M}. In that sense, FRB 20240114A does not appear to be an isolated case, but may belong to a subset of repeaters whose magneto-ionic environments evolve smoothly over month-to-year timescales rather than through purely stochastic fluctuations. At the same time, the detailed amplitudes, durations, and possible reversals of the RM evolution differ between sources, indicating diversity in the underlying environments and evolutionary drivers.

\subsection{Complex and Variable Emission Geometry}\label{subsec:emission}

Recent systematic studies of large samples of repeating FRBs observed with FAST have revealed that while some bursts exhibit PA swings consistent with the rotating vector model (RVM), the fitted geometric parameters and effective spin periods are highly inconsistent across bursts. This suggests that the magnetosphere of the FRB central engine is constantly distorted by the FRB emitter, leading to a dynamically evolving magnetic configuration rather than a stable dipole structure \citep{2025ApJ...988..175L}.

\begin{figure}
    \centering
    \includegraphics[width=0.9\textwidth]{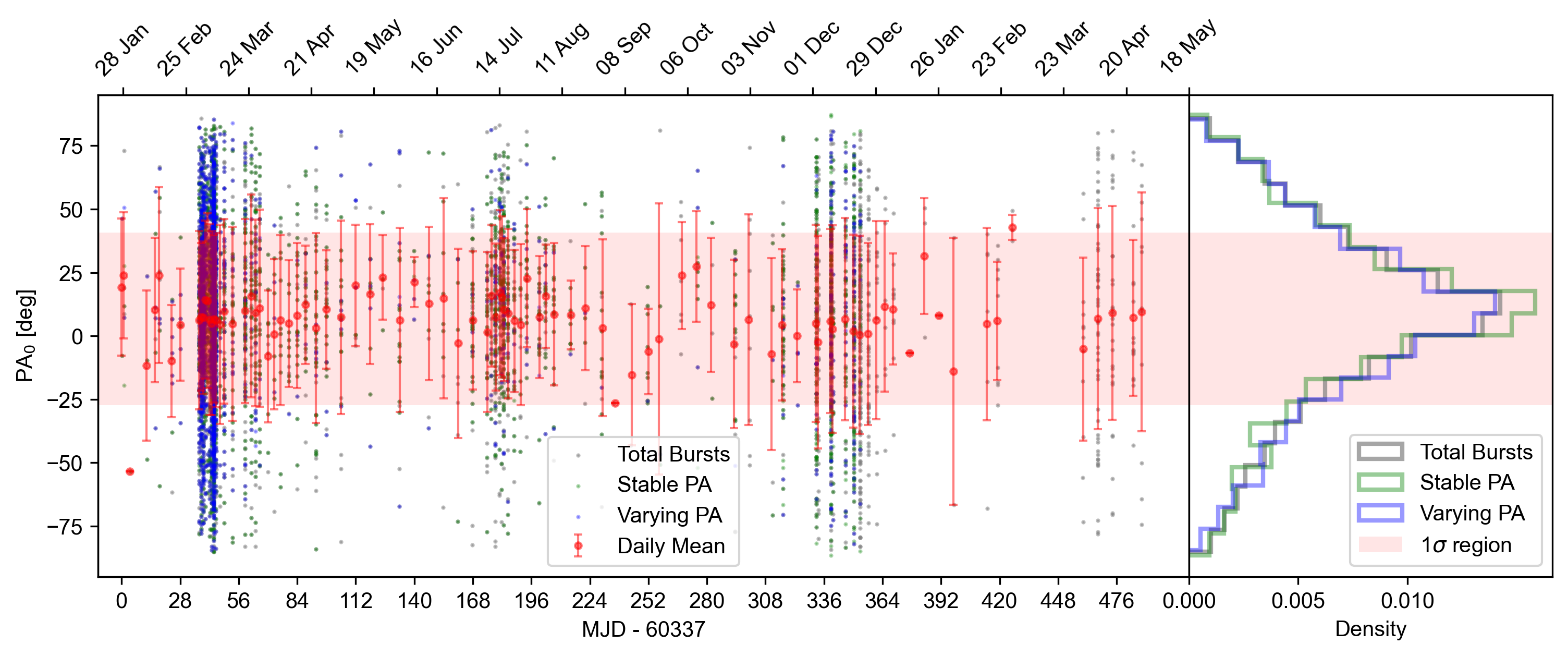}
    \caption{\textbf{Temporal evolution and distribution of the intrinsic polarization position angle (PA$_0$).} (Left) PA$_0$ as a function of time. Bursts are color-coded based on the reduced chi-square ($\chi_{\nu}^{2}$ ) of a constant-PA fit to their time-resolved, frequency-averaged PA measurements: green points indicate bursts with stable PA profiles ($\chi_{\nu}^{2} <$5), while blue points indicate bursts with variable PA profiles ($\chi_{\nu}^{2} \geq$5) \citep{2024ApJ...968...50P}. The red points with error bars represents the mean PA$_0$ and its standard deviation for each observing session. (Right) Histograms of the PA$_0$ distribution. The grey histogram includes all bursts, with the shaded region marking the $\pm$1$\sigma$ range. The green histogram includes only bursts with stable PA profiles (green points from the left panel) while the blue histogram includes bursts with variable PA profiles (blue points from the left panel). The red-shaded 1$\sigma$ region highlights the significant burst-to-burst PA variation present in the full sample.}
    \label{fig:PAevo}
\end{figure}

The polarization position angle evolution of FRB 20240114A also exhibits this dynamic characteristic. For each burst, we first determine its RM and derotate the Stokes $Q$ and $U$ data accordingly. We then construct a time-resolved PA profile from the frequency-summed derotated Stokes parameters. The uncertainty of each PA sample, denoted as PAE, is estimated by propagating the off-pulse noise in $Q$ and $U$; only on-pulse samples with PAE$\leq10^\circ$ are retained. We define PA$_0$ as the time average of the retained derotated PA samples for each burst. Unlike FRB 20180916B, which shows remarkably stable burst-to-burst PAs indicative of a highly stable magnetic field geometry at the emission site \citep{2025A&A.Bethapudi}, FRB 20240114A exhibits large burst-to-burst PA variation (Figure \ref{fig:PAevo}). Using the reduced-$\chi^2$ of a constant-PA fit to the derotated time-resolved PA profiles, we find that 67.2\% of bursts satisfy $\chi_\nu^2<5$. However, $\chi_\nu^2\geq5$ does not necessarily imply a prominent PA swing; if we additionally require the difference between the maximum and minimum PA within a burst to exceed 40$^\circ$, the fraction of bursts with clear intraburst PA swings is 17.6\%. For bursts with variable PA profiles, PA$_0$ should therefore be regarded as a representative value rather than a complete description of the full intraburst PA evolution. The broad distribution of PA$_0$, even for bursts with stable intra-burst PA profiles (blue histogram in Figure \ref{fig:PAevo}), and the significant scatter of the session-averaged PA$_0$ (red error bars in Figure \ref{fig:PAevo}, left) suggest that the bursts originate from a spatially complex or temporally fast-evolving magnetospheric environment. The lack of a stable, globally preferred PA challenges models where emission is strictly confined to a fixed magnetic pole. The combination of a systematically evolving RM (probing the large-scale, intervening screen) and a highly variable PA (probing the immediate emission geometry) suggests that both the large-scale environment and the small-scale emission region are dynamic, albeit likely on different timescales.

Interestingly, independent observations of FRB~20240114A with the Parkes ultra-wideband low (UWL) receiver have revealed a multi-peak distribution in the polarization position angle, with excesses near -17.39 degree and 64.8 degree \citep{2026arXiv260216409U}. This contrasts with the PA distribution observed in our FAST data, suggesting that the apparent PA distribution may depend on the observed frequency band or the specific activity epoch. Such frequency-dependent PA behavior would be expected in a dynamically evolving magnetosphere, where different emission heights or regions could be probed at different frequencies. However, the effect appears to be relatively weak, as both datasets consistently point to a highly variable emission geometry without a single preferred direction.

\subsection{Comparison with Other Repeaters and the Origin of Periodicity}

The possible quasi-periodic modulation of burst activity in FRB~20240114A reported by \cite{2026arXiv260216409U}, introduces an additional layer of complexity to the understanding of this source. Using Murriyang observations, they proposed a timescale of approximately 70 days and interpreted it as evidence for plasma lensing due to extreme scattering events.  \cite{2025arXiv250714708Z} have also identified periodicities in FAST observations of this source, with periods of $143.40 \pm 7.19$ days (5.2$\sigma$ significance) and $73.60 \pm 2.45$ days (3.3$\sigma$ significance). Notably, periodic phenomena are not unique to FRB~20240114A. Several well-studied repeating FRBs have been reported to exhibit periodicities in either their burst activity or environmental properties, including FRB~20180916B, which shows a 16.35-day activity period \citep{2020Natur.582..351C}. In this context, any potential activity periodicity in FRB~20240114A would place it among a small but growing group of repeaters with long-term activity modulation. The diversity of reported activity-modulation timescales among repeaters suggests that multiple physical mechanisms may be at play.
The interpretation of any potential periodicity in FRB~20240114A thus remains an open question, necessitating further monitoring and multi-wavelength observations to distinguish between intrinsic and environmental origins.

\section{Summary and Data Access}\label{sec:summary}

In this paper, we present a large-scale polarization catalog for the repeating FRB~20240114A, resulting from an extensive monitoring campaign with the FAST telescope between 2024 January 28 and 2025 May 30. The core of this work is the detection and parameterization of 6,131 bursts. We have detailed the comprehensive data processing pipeline, from initial data acquisition and pre-processing to sophisticated polarization calibration and RM synthesis. The key deliverables of this pipeline are precise measurements for each burst, including the topocentric arrival time (MJD), DM, pulse width, bandwidth (BW), RM, and the degrees of linear and circular polarization.

The global analysis of this catalog reveals several key characteristics of the source:

\begin{itemize}
    \item \textbf{Dynamic Rotation Measure}: The RM exhibits a remarkable two-phase temporal evolution: an initial period of relative stability around a mean value, followed by a rapid, near-linear decay of approximately 200~rad~m$^{-2}$ over about 200 days. This evolution manifests as a bimodal distribution in the RM histogram. The joint RM–DM temporal analysis indicates that the RM variation is dominated by changes in the line-of-sight magnetic field strength rather than electron column density, implying a dynamically evolving magnetized environment in the immediate vicinity of the source.
    \item \textbf{Stable Dispersion Measure}: In contrast to the RM, the DM remains remarkably stable throughout the entire observing campaign, indicating a constant integrated electron density along the line-of-sight.
    \item \textbf{High Degree of Polarization}: The linear polarization fraction is generally high, while the circular polarization is typically low. The relationship between linear and circular polarization for the vast majority of bursts adheres to the theoretical physical limit, with outliers primarily attributed to low-SNR measurements. The intrinsic polarization position angle distribution is broad and centered near 10$^\circ$ with a non-Gaussian excess, but shows no stable, singular preferred orientation over time.
    \item \textbf{Stable Polarization Fraction Distributions}: Despite the dramatic evolution in RM, the combined two-dimensional distribution of linear versus circular polarization fractions remains statistically stable across the entire observing period. This suggests that the underlying emission mechanism and the net ordered magnetic field geometry in the immediate emission region are largely invariant.
    \item \textbf{Absence of Frequency-Dependent Depolarization}: Analysis of bright, broadband bursts shows no systematic decrease of linear polarization fraction with frequency, arguing against significant random magnetic fields ($\sigma_{\text{RM}}$) within the main Faraday rotating screen.
    \item \textbf{Absence of Faraday Conversion}: No significant Faraday conversion is detected at any epoch. This non-detection indicates that the polarization data probe only the line-of-sight magnetic field component. The perpendicular component remains unconstrained. Future ultra–wideband and higher-sensitivity observations may detect conversion. This would enable joint tracking of the perpendicular and parallel magnetic field components, and allow reconstruction of the time-varying three-dimensional magnetic field in the source vicinity.
\end{itemize}

The global distributions and correlations of these parameters, presented in Section \ref{sec:quality}, attest to the high quality and internal consistency of the catalog. They reveal a source with a stable DM but a highly dynamic RM, complex polarization behavior, and no simple correlation between DM and RM. These features underscore the value of this uniform, high-sensitivity dataset for statistical and temporal evolution studies.

The complete catalog is publicly available to the scientific community to maximize its utility. The catalog, in CSV formats, can be accessed through the following permanent repositories: \url{10.57760/sciencedb.Fastro.00040}. The original observational data are available from the FAST archive: \url{https://fast.bao.ac.cn/}.

The catalog is also available as a machine-readable table in the electronic edition of this journal. We encourage the community to use this catalog for further investigations.

\begin{acknowledgments}

We thank the staff of the FAST telescope for their support during the observations. FAST is a Chinese national mega-science facility, built and operated by the National Astronomical Observatories, Chinese Academy of Sciences. This work made use of the data from FAST FRB Key Science Project and the data of the CHIME/FRB VOEvent Service, BlinkVerse and TransientVerse.

This study is supported by National Natural Science Foundation of China (NSFC) Programs Nos. 12588202, 11690024, 11725313, 11988101, 12041303, 12203045, 12233002, 12303042, 12403100, 12421003, 12447115, 12522305, 12541304, 12373040, U1731238, U2031117, W2442001; the CAS International Partnership Program No. 114-A11KYSB20160008; the CAS Strategic Priority Research Program No. XDB23000000; the National Key R\&D Program of China Nos. 2021YFA0718500, 2023YFE0110500, QN2023061004L, 2024YFA1611703, 2024YFA1611700, the National SKA Program of China No. 2022SKA0130104, the China Postdoctoral Science Foundation (CPSF) under Grant Nos. GZB20240308, GZB20250737, 2025T180875 and the Fundamental Research Funds for the Central Universities. 
P.W. acknowledges support from the CAS Youth Interdisciplinary Team, the Youth Innovation Promotion Association CAS (id. 2021055), and the Cultivation Project for FAST Scientific Payoff and Research Achievement of CAMS-CAS.
D.L. is a New Cornerstone investigator.
W.W.Z. is supported by the CAS Project for Young Scientists in Basic Research, YSBR-063. 
Y.F.H. is supported by the Xinjiang Tianchi Program.
Y.F. is supported by the Leading Innovation and Entrepreneurship Team of Zhejiang Province of China grant No. 2023R01008.
W.-Y. W. acknowledges support from the NSFC (No.12261141690 and
No.12403058), and the Strategic Priority Research Program of the CAS (No. XDB0550300).
J. R. Niu is supported by the National Natural Science Foundation of China (NSFC, No. 12503055) and the Postdoctoral Fellowship Program of CPSF under Grant Number GZB20250737.
R. S. Zhao is supported by the National Natural Science Foundation of China (Grant Nos. 12563008), Science and Technology Foundation of Guizhou Provincial Department of Education (No. KY(2023)059), Liupanshui Science and Technology Development Project (No. 52020-2024-PT-01). 
Q. Wu is supported by the National Natural Science Foundation of China (Grant Nos. 12447115 and 12503050) and the China Postdoctoral Science Foundation (CPSF) under Grant Number GZB20240308, 2025T180875, and 2025M773199.

\end{acknowledgments}

\begin{contribution}

All authors discussed the contents and form the final version of the paper.


\end{contribution}

\facility{FAST}

\software{astropy \citep{2013AstropyCollaborationA&A},  
          SciPy \citep{2020VirtanenNatMe}, 
          DM-power \citep{2022Lin},
          PRESTO \citep{2001AAS.Ransom},
          DRAFTS \citep{2025ApJS.ZhangYK},
          PSRCHIVE \citep{2004PASA.Hotan},
          ionFR \citep{2013ascl.soft03022S}
          }

\appendix

\section{Polarization Calibration}\label{sec:PolCal}

The polarization of the detected bursts was calibrated by correcting for differential gains and phases, using separate measurements of a noise diode injected at an angle of 45$^\circ$ from the linear receiver. This was done with the software package PSRCHIVE \citep{2004PASA.Hotan}. The circular polarization of most bursts is consistent with noise, being less than a few percent of the total intensity, in agreement with the findings of \cite{2018Natur.Michilli}.

\begin{figure}
    \centering
    \includegraphics[width=0.9\textwidth]{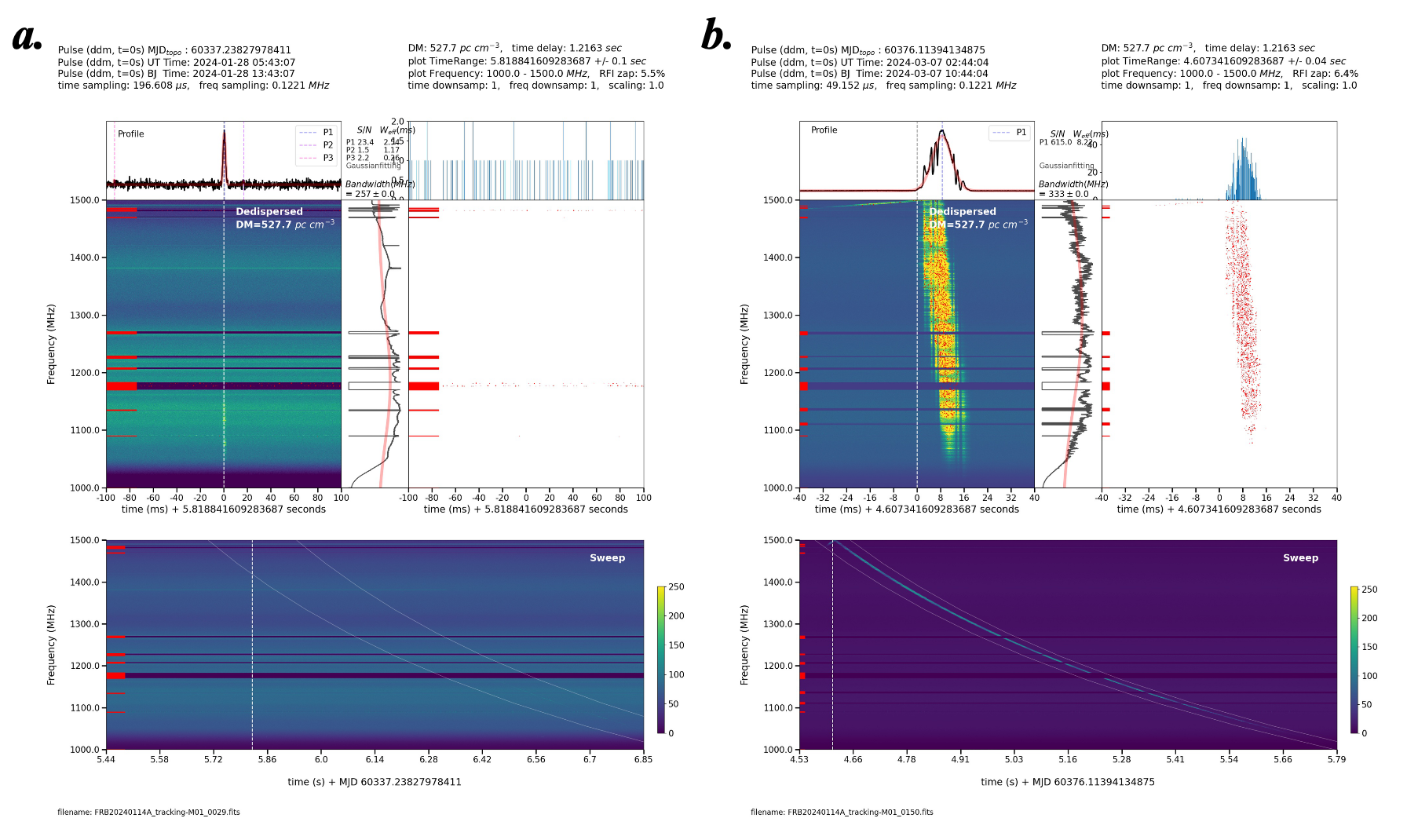}
    \caption{\textbf{Examples of dynamic spectra (waterfall plots) illustrating the identification of anomalous samples.} In both cases, radio frequency interference (RFI) and samples affected by signal saturation are marked in red. (a) A burst without saturation. The red points are predominantly confined to specific, persistent RFI frequency channels. Consequently, the time profile of these flagged samples shows a relatively uniform distribution, reflecting the continuous nature of the RFI. (b) A burst with saturation. The red points are concentrated within the bright, central region of the burst itself, indicating that the high signal strength has driven the digitizer or pipeline into a non-linear regime. This saturation leads to a distinct peak in the time profile of the flagged samples, precisely coinciding with the burst location. These examples validate the effectiveness of our automated flagging routines in distinguishing between general RFI and instrumental saturation, ensuring the integrity of the burst profiles and measured parameters used in the catalog.}
    \label{fig:saturation}
\end{figure}

Saturation occurs in the FAST receiver when radio flux densities reach levels between hundreds of Jansky and mega-Jansky. To identify potential saturation events and mitigate their impact on polarization analysis, each burst's four polarization components (XX, YY, X$^*$Y, Y$^*$X) were systematically examined. Here the quoted thresholds refer to the raw stored intensity values in the FAST FITS files rather than calibrated physical units. Saturation manifests differently across components: XX and YY saturate when intensities exceed 255 (wrapping to 0), while X$^*$Y and Y$^*$X saturate when absolute intensities surpass 127 (also wrapping to 0). For every burst, the dynamic spectra of all four polarization components were inspected. Critical-value data points (corresponding to the 255 and 127 thresholds) were flagged on the total intensity (I) dynamic spectra. These points were then frequency-integrated to generate time profiles. Two distinct scenarios were identified:
Time profiles showing uniform temporal distribution indicate noise or radio frequency interference (RFI) effects (see Fig. \ref{fig:saturation}$a$).
Time profiles matching the burst temporal structure signify genuine saturation (see Fig. \ref{fig:saturation}$b$).
Bursts exhibiting the latter characteristic were excluded from subsequent analysis. It should be noted that the value 255 or 127 itself does not directly indicate saturation. The key criterion for identifying saturation is observing whether the distribution of tracers aligns with the pulse morphology. However, this serves only as an operational guideline, as the underlying principles are quite complex. Consequently, in practice there is no single unique method for determining saturation. Applying this methodology to the sample of bursts with SNR$\geq$20 revealed 253 saturated events, which were subsequently removed to ensure reliable polarization measurements.

\section{RM Synthesis}\label{sec:RMsyn}

Due to the Faraday effect, the polarization plane of the FRB signal rotates during propagation. We employed the RM synthesis method to search for the RM of the burst by considering the equation of
\begin{equation}\label{Eq:IQUV}
    \begin{pmatrix}
        I \\
        Q^{\prime} \\
        U^{\prime} \\
        V
    \end{pmatrix}
    =
    \begin{pmatrix}
        1 & 0 & 0 & 0 \\
        0 & \cos\ 2\theta & \sin\ 2\theta & 0 \\
        0 & -\sin\ 2\theta & \cos\ 2\theta & 0 \\
        0 & 0 & 0 & 1 
    \end{pmatrix}
    \begin{pmatrix}
        I \\
        Q \\
        U \\
        V
    \end{pmatrix},
\end{equation}
 where $ \theta = \text{RM}\lambda^2 $. In our implementation, the trial RM range is set to -2000 to +2000 rad~m$^{-2}$. This interval was chosen as a conservative search range, substantially broader than the RMs measured for FRB~20240114A, while avoiding an unnecessarily wide search domain. The intrachannel depolarization factor can be written as $f_\mathrm{depol}=|\sin(\mathrm{RM}\Delta\lambda^2)/(\mathrm{RM}\Delta\lambda^2)|$, where $\Delta\lambda^2\approx2c^2\Delta\nu/\nu_c^3$ in the narrow-channel approximation \citep{1966MNRAS.133...67B,2005A&A...441.1217B}. For the present FAST L-band channelization ($\Delta\nu\approx$0.122 MHz), 10\% depolarization at the bottom of the band occurs only at $|\mathrm{RM}|\sim1.8\times10^4$ rad~m$^{-2}$, far above both our adopted search range and the RMs measured in this work. The RM uncertainty is estimated empirically from the width of the RM-synthesis peak. Specifically, after identifying the RM corresponding to the maximum of the RM spectrum, we determine the left and right peak widths using the points where the spectrum drops below the midpoint between its maximum and minimum values, and divide these widths by the burst S/N. This yields asymmetric RM uncertainties when the RM-synthesis peak is not perfectly symmetric. For the present catalog, the RM analysis is restricted to bursts with S/N $\geq$ 20, for which the RM-synthesis peak is in most cases visually well defined. In practice, the corresponding polarization and RM-synthesis diagnostic plots were also inspected to identify obviously spurious cases. A more formal calibration of peak significance could in principle be obtained through dedicated Monte Carlo simulations, but this is beyond the scope of the present catalog paper. In addition to the empirical peak-width-based RM uncertainty quoted here, one RM-search grid step may be regarded as an additional systematic uncertainty ($\sim$ 0.2 rad~m$^{-2}$) associated with the discretization of the RM search. Due to challenges in calculating RM from weak signals or bursts with narrow emission bandwidths, RM values were successfully measured only 6,131 of the detected bursts. The mean value of the RM is 359.52 $\rm rad\ m^{-2}$, showing significant variations throughout the observing campaign, ranging from 212.50 $\rm rad\ m^{-2}$ to 420.50 $\rm rad\ m^{-2}$. We corrected the Stokes component $Q$ and $U$ by derotating it using the measured RM. The degree of linear and circular polarization are calculated as,
\begin{equation}\label{Eq:DoP}
    L/I = \frac{\Sigma_i L_i}{\Sigma_i I_i},\ V/I = \frac{\Sigma_i V_i}{\Sigma_i I_i},
\end{equation}
where $I_i$ and $V_i$ are defined as the frequency-averaged total intensity and circular polarization measured for the time sample $i$, $L_i = \sqrt{Q_i^2 + U_i^2}$ is the frequency-averaged total linear polarization measured for the time sample $i$ , respectively. Due to the quadrature summation $ L = \sqrt{Q^2+U^2} $, the probability density function of $L$ is non-Gaussian (see below). In addition, there exists a positive bias in the measured values at low $L$, which means the linear polarization can be subject to overestimation when noise is present. To remove the bias and calculate the true value of linear polarization, we employ the following equation to de-bias \citep{2001Everett}:

\begin{equation}\label{Eq:L_de-bias}
    L_{\mathrm{de-bias}} = \begin{cases}
        \sigma_I \sqrt{(\frac{L_i}{\sigma_I})^2-1}, & \frac{L_i}{\sigma_I} > 1.57 \\
        0, & \frac{L_i}{\sigma_I} < 1.57
    \end{cases} \ \ \ ,
\end{equation}
where $\sigma_I$ is the off-pulse standard deviation in Stokes parameter $I$.

For a subset of low-S/N bursts, the formally measured $L/I$ can exceed 100\% (see Figure~\ref{fig:DOP-SNR}). These cases are preferentially associated with bursts whose intrinsic linear polarization is already very high, i.e. close to complete linear polarization. In such events, off-pulse noise fluctuations can occasionally yield $L>I$ in the background, and when the burst itself has relatively low integrated S/N, the accumulated result can produce $\sum L > \sum I$. We therefore interpret these values not as physically meaningful super-100\% polarization fractions, but as noise-biased measurements of bursts whose true $L/I$ is likely close to 100\%. 

\section{Correction for Ionospheric Faraday Rotation}\label{sec:IonCorr}

\begin{figure}
    \centering
    \includegraphics[width=0.9\textwidth]{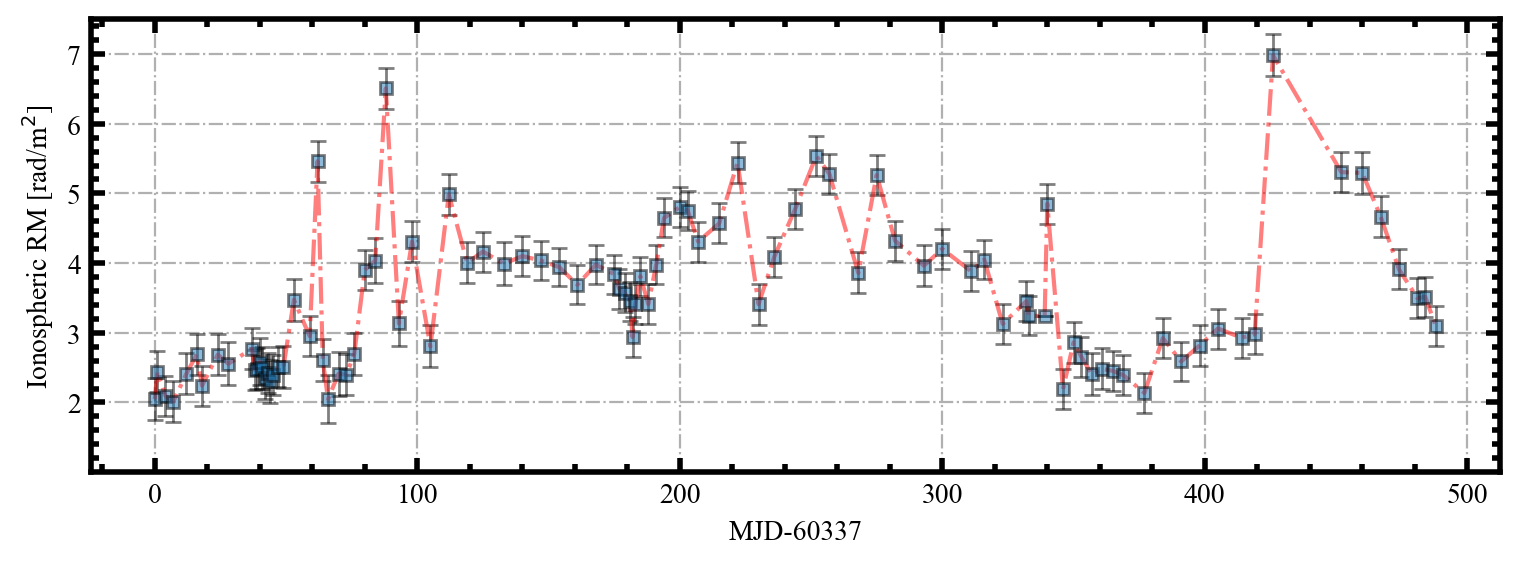}
    \caption{\textbf{Contribution of the Earth’s ionosphere to the observed rotation measure.} The plot shows the ionospheric RM as a function of time. The mean ionospheric RM throughout the campaign is 3.49 $\pm$ 1.06 $rad\ m^{-2}$.}
    \label{fig:IonRM}
\end{figure}

The Earth ionosphere contributes an additive, time-variable component to the observed RM. To quantify and remove this effect, we calculate the ionospheric RM for each observing session using a Python code IonFR and global ionospheric map (GIM) products. The GIM products are available on NASA’s website\footnote{\url{cddis.nasa.gov/archive/gnss/products/ionex/}} in the IONosphere EXchange (IONEX) format. The ionospheric RM as a function of time is presented in Figure \ref{fig:IonRM}. Throughout our entire observing campaign, the ionospheric RM has a mean value of 3.49 $\rm rad\ m^{-2}$ with a standard deviation of 1.06 $\rm rad\ m^{-2}$. This contribution is over an order of magnitude smaller than the $\sim$200 $\rm rad\ m^{-2}$ intrinsic RM evolution we report for FRB 20240114A. More importantly, its temporal variability is negligible compared to the source intrinsic changes. Therefore, we conclude that the ionospheric contribution does not account for the observed dramatic RM evolution. All RM values presented in the final catalog have been corrected by subtracting the ionospheric RM for the corresponding MJD.

\section{Energy Distance}\label{sec:ED}

\textbf{Energy distance} is a statistical metric that quantifies 
the discrepancy between two multi-variate probability distributions. 
It operates directly on the moments of inter-point distances, 
offering a non-parametric and geometrically intuitive measure of 
distributional difference \citep{2004Szekely,SZEKELY2013EnergyStatistics}.

\textbf{Definition.} Let \( X \) and \( X' \) be independent and 
identically distributed (i.i.d.) random vectors from distribution 
\( P \), and let \( Y \) and \( Y' \) be i.i.d. from distribution 
\( Q \), all defined on \( \mathbb{R}^d \) with finite first moments. 
The energy distance \( \mathcal{D}_E(P, Q) \) is defined as 
\citep{2004Szekely,SZEKELY2013EnergyStatistics}:
\[
\mathcal{D}_E(P, Q) = 2 \, \mathbb{E} \| X - Y \| 
- \mathbb{E} \| X - X' \| - \mathbb{E} \| Y - Y' \|,
\]
where \( \mathbb{E} \) denotes the expectation and \( \|\cdot\| \) 
is the Euclidean norm (or, more generally, a suitable metric).

\textbf{Key Properties.} The energy distance satisfies the following 
essential properties \citep{2004Szekely,SZEKELY2013EnergyStatistics,SZEKELY2013DistanceCorrelation}:

1. \textbf{Non-negativity \& Identity:} \( \mathcal{D}_E(P, Q) \geq 0 \), 
and \( \mathcal{D}_E(P, Q) = 0 \) if and only if \( P = Q \) almost 
everywhere. This makes it a proper statistical \emph{distance} or \emph{metric}.

2. \textbf{Equivalence via Characteristic Functions:} An alternative, 
powerful characterization links it to the characteristic functions 
\( \phi_P(t) \) and \( \phi_Q(t) \) of the distributions \citep{2004Szekely,SZEKELY2013EnergyStatistics}:
\[
\mathcal{D}_E(P, Q) = \frac{1}{c_d} 
\int_{\mathbb{R}^d} \frac{ | \phi_P(t) - \phi_Q(t) |^2 }{ \| t \|^{d+1} } dt,
\]
where \( c_d \) is a dimensional constant. This formulation shows that 
\( \mathcal{D}_E \) measures the weighted \( L^2 \)-difference between 
the characteristic functions.

3. \textbf{Sample Estimator:} Given empirical samples 
\( \{x_i\}_{i=1}^n \sim P \) and \( \{y_j\}_{j=1}^m \sim Q \), 
\( \mathcal{D}_E \) can be estimated unbiasedly from inter-point distances, 
making it applicable to real-world data \citep{SZEKELY2013EnergyStatistics}.

\textbf{Applications.} In statistical and machine learning contexts, 
energy distance is widely used for two-sample hypothesis testing 
(the \emph{energy test}), assessing the convergence of generative models, 
and as a loss function in domain adaptation 
\citep{2004Szekely,SZEKELY2013EnergyStatistics,sejdinovic2013Sejdinovic}. 
Its primary advantages are its metric properties, sensitivity to all 
moments, and computational feasibility for moderate sample sizes.

\textbf{Note.} In the main text, we use the energy distance to quantify the difference between the joint distributions of the linear and circular polarization fractions of cumulative bursts in consecutive observing sessions over different time-scales. Specifically, each burst is represented as a two-dimensional vector,
\begin{equation}
\mathbf{x}_i = \left(\frac{L_i}{I_i},\, \frac{V_i}{I_i}\right),
\end{equation}
where $L_i/I_i$ and $V_i/I_i$ denote the linear and circular polarization fractions, respectively. For two samples $X=\{\mathbf{x}_i\}_{i=1}^{m}$ and $Y=\{\mathbf{y}_j\}_{j=1}^{n}$ from two observing sessions, we compute the sample energy distance as
\begin{equation}
\mathcal{E}(X,Y)
=
\frac{2}{mn}\sum_{i=1}^{m}\sum_{j=1}^{n}\|\mathbf{x}_i-\mathbf{y}_j\|
-
\frac{1}{m^{2}}\sum_{i=1}^{m}\sum_{j=1}^{m}\|\mathbf{x}_i-\mathbf{x}_j\|
-
\frac{1}{n^{2}}\sum_{i=1}^{n}\sum_{j=1}^{n}\|\mathbf{y}_i-\mathbf{y}_j\|,
\end{equation}
where $\|\cdot\|$ is the Euclidean distance in the two-dimensional polarization-fraction space. A larger value of $\mathcal{E}$ indicates a greater discrepancy between the two joint distributions.




\bibliography{sample701}{}
\bibliographystyle{aasjournalv7}



\end{document}